\documentclass[12pt,draftclsnofoot,onecolumn]{IEEEtran}
\usepackage{amsmath,graphics,amssymb,epsfig,subfigure,color,amsthm,cite,float}
\usepackage{array}
\usepackage{multirow}
\usepackage{enumerate}
\usepackage{makecell}
\usepackage{tabularx} 
\usepackage{algorithm,algcompatible}
\algnewcommand\INPUT{\item[\textbf{Input:}]}%
\algnewcommand\OUTPUT{\item[\textbf{Output:}]}%

\newtheorem{lem}{Lemma}

\begin{document}
\title{Spatial Lattice Modulation for MIMO Systems}
\author{Jiwook Choi,
        Yunseo Nam,
        and~Namyoon Lee
        \thanks{J. Choi, Y. Nam, and N. Lee are with the department of electrical engineering, POSTECH, South Korea, emails:\{jiwook,edwin624,nylee\}@postech.ac.kr
}
\thanks{}}
\markboth{}%
{Shell \MakeLowercase{\textit{et al.}}: Bare Demo of IEEEtran.cls for IEEE Journals}
\maketitle
\begin{abstract}

This paper proposes spatial lattice modulation (SLM), a spatial modulation method for multiple-input-multiple-output (MIMO) systems. The key idea of SLM is to jointly exploit spatial, in-phase, and quadrature dimensions to modulate information bits into a multi-dimensional signal set that consists of lattice points. One major finding is that SLM achieves a higher spectral efficiency than the existing spatial modulation and spatial multiplexing methods for the MIMO channel under the constraint of $M$-ary pulse-amplitude-modulation (PAM) input signaling per dimension. In particular, it is shown that when the SLM signal set is constructed by using dense lattices, a significant signal-to-noise-ratio (SNR) gain, i.e., a nominal coding gain, is attainable compared to the existing methods. In addition, closed-form expressions for both the average mutual information and average symbol-vector-error-probability (ASVEP) of generic SLM are derived under Rayleigh-fading environments. To reduce detection complexity, a low-complexity detection method for SLM, which is referred to as \textit{lattice sphere decoding}, is developed by exploiting lattice theory. Simulation results verify the accuracy of the conducted analysis and demonstrate that the proposed SLM techniques achieve higher average mutual information and lower ASVEP than do existing methods.

\end{abstract}
\IEEEpeerreviewmaketitle
\section{Introduction}
 Spatial modulation (SM) \cite{Mesleh2008} is a transmission method that sends information bits using the index of an active antenna and conventional quadrature-amplitude-modulation (QAM) symbols. SM has been proposed to improve both the spectral and energy efficiency of MIMO systems \cite{Renzo2011,Renzo2014,Yang2015,Yang2016}. For example, when a transmitter is equipped with $N_{\rm t}$ antennas that use a radio frequency (RF) chain, $ \log_2{(N_{\rm t})} +  \log_2(|\mathcal{C}_{\rm Q}|)$ bits can be modulated into a spatial symbol vector, where $ \mathcal{C}_{\rm Q}$ denotes the QAM constellation set and $|\mathcal{C}_{\rm Q}|$ represents its cardinality. A simplified version of SM, referred to as space shift keying (SSK) \cite{Jeganathan2009} was presented to improve energy efficiency. SSK only maps information bits into the antenna index, so it is able to achieve the spectral efficiency of $\log_2{(N_{\rm t})}$ bits/sec/Hz when signal-to-noise ratio (SNR) is high enough. 
	
The concepts of SM and SSK have been generalized in numerous ways by mapping information bits into multiple indices of the transmit antennas. Generalized spatial modulation (GSM) \cite{Younis2010} and generalized space shift keying (GSSK) \cite{Jeganathan2010} are representative generalizations of SM. The idea of both GSM and GSSK is to map information bits onto an antenna subset that consists of $N_{\rm a}$ elements among $N_{\rm t}$. Therefore, a transmitter is able to modulate $ \log_2{{N_{\rm t} \choose N_{\rm a}}} +  \log_2(|\mathcal{C}_{\rm Q}|)$ information bits when using GSM with constellation set $\mathcal{C}_{\rm Q}$. This modulation strategy allows sending of $ \log_2{{N_{\rm t} \choose N_{\rm a}}} -  \log_2{{N_{\rm t} \choose 1}} $ more information bits than both SM and SSK. Multiple active spatial modulation (MA-SM)\cite{Wang2012} is another variation of GSM, MA-SM was introduced by harnessing multiplexing gains of the MIMO system. MA-SM sends distinct QAM symbols by choosing $N_{\rm a}$ active elements among $N_{\rm t}$ transmit antennas; thereby, $ \log_2{{N_{\rm t} \choose N_{\rm a}}} +  N_{\rm a}\log_2(|\mathcal{C}_{\rm Q}|) $ information bits are modulated to a symbol vector with QAM constellation set $\mathcal{C}_{\rm Q}$. 

Variable set of active antenna GSM (VA-GSM) \cite{Humadi2015} is another variation of GSM. VA-GSM allows the number of active antennas to vary from $1$ to $N_ {\rm t}$, while sending the same transmit symbol for each active antenna. In addition, quadrature spatial modulation (QSM) \cite{Mesleh2015} separately exploits in-phase and quadrature signal dimensions. For instance, when a transmitter has $N_{\rm t}$ antennas, QSM is able to modulate $2\log_2\left(N_{\rm t}\right)+\log_2\left(|\mathcal{C}_{\rm Q}|\right)$ information bits. Recently, another generalized method of GSM, called GSM with multiplexing (GSMM) \cite{Ibrahim2016}, was introduced. GSMM sends $S$ data symbols using a set of precoding matrices $\mathcal{F}$ in which $1\leq S\leq N_{\rm t}$ and $|\mathcal{F}|=2^{N_{\rm t}}-1$. As a result, GSMM is able to modulate $\log_2\left(2^{N_{\rm t}}-1\right) + S\log_{2}(|\mathcal{C}_{\rm Q}|)$ information bits. The common limitation of the methods in \cite{Younis2010, Jeganathan2010,Wang2012,Humadi2015,Mesleh2015,Ibrahim2016} is that information bits are separately modulated to the index of active antenna-subsets (or the index of precoding matrices \cite{Ibrahim2016}) and to the transmission of QAM symbols. This separated modulation approach generally cannot achieve a higher spectral efficiency than that attained by a joint modulation strategy for a fixed $N_{\rm t}$ and $\mathcal{C}_{\rm Q}$. 

Adaptive-joint-mapping GSM (AJM-GSM) \cite{Ma2012} and jointly-mapped SM (JM-SM) \cite{Guo2016} both jointly modulate information bits into an active antenna index and transmit symbols using the proposed joint mapping rule. This joint mapping can generate more signal points than the mapping methods that separately modulate information bits into the indices of antenna subsets and transmit symbols. The limitation of the methods in \cite{Mesleh2015} and \cite{Guo2016} is that they do not jointly take into account all possible signaling dimensions, (i.e., spatial, in-phase, quadrature) when constructing signal sets for SM.


Since the minimum Euclidean distance between adjacent symbols in a constellation set affects the detection accuracy, numerous methods for the signal set design of SM have been proposed to maximize the minimum distance \cite{Cheng2015,Cheng2016,Freudenberger2017}. For instance, the QAM symbols of MA-SM \cite{Wang2012} are rotated to a different phase for each active antenna-subset; as a result, the minimum distance increases. Enhanced spatial modulation (ESM) \cite{Cheng2015,Cheng2016} uses a similar approach. The idea of ESM is to construct a primary and a secondary signal constellation set. Then the secondary set is interpolated into the primary set to increase the minimum distance between symbol vectors. Another constellation design method for SM exploits Eisenstein integer \cite{Freudenberger2017}; in this method, a transmitter uses a two-dimensional hexagonal lattice to send symbols. This lattice is the densest packing lattice in a two-dimensional complex domain. All these methods \cite{Cheng2015,Cheng2016,Freudenberger2017} demonstrate that carefully-designed constellation sets is able to achieve higher SNR gains than SM methods that use the conventional QAM constellation set. 

In this paper, we consider a MIMO system in which a transmitter is equipped with $N_{\rm t}$ transmit antennas and a receiver is equipped with $N_{\rm r}$ receive antennas. We assume that the transmitter can send one symbol per in-phase (or quadrature) component of each transmit antenna from the $M$-ary pulse-amplitude-modulation (PAM) signal set $\mathcal{C}=\left\{\frac{-M}{2},\frac{-M}{2}+1,\ldots,\frac{M}{2}-1,\frac{M}{2}\right\}$. The contributions of this paper are summarized as follows:

\begin{itemize}
	\item Our major contribution is to propose a novel multidimensional spatial modulation method called \textit{spatial lattice modulation (SLM)}. Unlike the existing SM techniques in which information bits are separately mapped to a set of antenna indices and modulation symbols \cite{Younis2010, Jeganathan2010,Wang2012,Humadi2015,Ibrahim2016}, the key idea of the proposed SLM is to modulate information bits into a set of lattice points in $\mathbb{R}^{2N_{\rm t}}$ by jointly exploiting spatial, in-phase, and quadrature signal dimensions. An element of each lattice point is chosen from the set $\mathcal{\bar C}=\{0\} \cup\mathcal{C}$ with $|\mathcal{\bar C}|=M+1$, where the null element $\{0\}$ indicates that the input signal for a chosen signal dimension is deactivated for SM. In particular, we present two SLM methods: SLM using a simple cubic lattice and SLM that uses a dense packing lattice that has a large nominal coding gain in a low-dimensional vector space \cite{Forney1998, ConwayBook}. We show that the proposed SLM methods achieve the spectral efficiency of $ 2N_{\rm t}\log_2\left( M+1\right)$ bits/sec/Hz when SNR is high enough. This result is interesting because it enables the transmission of $2N_{\rm t}\left( \log_2\left( M+1\right) -  \log_2\left( M\right)\right)$ additional information bits per channel use compared to the conventional spatial multiplexing; this gain is unbounded as $N_{\rm t}$ increases. In particular, to attain a nominal coding gain for a given target spectral efficiency, we propose a signal set design algorithm that uses Barnes-Wall lattices for SLM; these are the densest lattices below 16 dimensions (or eight transmit antennas).

\item We also analyze the average mutual information and average symbol-vector-error-probability (ASVEP) of the proposed SLM methods under a Rayleigh MIMO channel. Although the mutual information expressions have been characterized for SM and GSM in \cite{Guan2013,Jin2015,Guo2014} for a fixed MIMO channel, the average mutual information expression is unknown for general SM methods. We derive a tight approximation of the average mutual information in a closed-form for the proposed SLM methods. We also derive a closed-form upper bound of ASVEP for SLM to complete our analysis. Simulation results verify the effectiveness of our analysis. One major observation is that SLM using dense lattices provides SNR gains in both the average mutual information and the ASVEP.

\item Lastly, we present a low-complexity SLM detection method, which is called \textit{lattice sphere decoding (LSD)}. The key idea of LSD is to exploit the property that a lattice is closed under addition. Using this lattice property, the proposed LSD algorithm reduces the effective search space of SLM by calculating ML metrics only in the closest lattice vectors from an initially estimated lattice vector.  We show that the complexity order of the proposed LSD algorithm is $\mathcal{O} \left( N_{\rm t}^{3}\right)$, which is the same detection complexity order with linear-type detection methods such as zero-forcing MIMO detection. Simulation results show that the error performance of LSD closely matches that of ML detection in practical MIMO systems, while significantly diminishing the detection complexity.

 
 
 \end{itemize}

\section{System Model and Preliminaryies}
\begin{figure*} [t]
	\centering
  \includegraphics[width=1\textwidth]{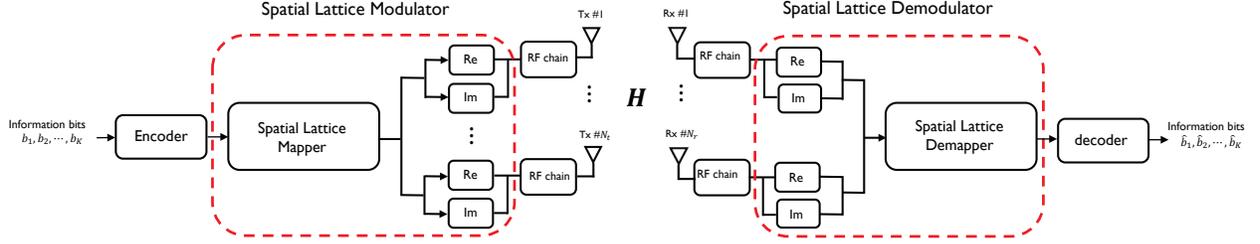}\\
  \caption{System architecture for  the proposed SLM in a MIMO system.} \label{SLM_tx}
\end{figure*}
This section presents the system model considered in this paper and provides some useful mathematical definitions that will be used subsequently.
\subsection{System Model}
We consider a MIMO channel in which a transmitter equipped with $N_{\rm t}$ transmit antennas sends information symbols to a receiver equipped with $N_{\rm r}$ receive antennas. We denote the complex baseband transmit vector by ${\bf \bar x} \in\mathbb{C}^{N_{\rm t}\times 1}$, and the MIMO channel matrix by ${\bf \bar H}\in\mathbb{C}^{N_{\rm r}\times N_{\rm t}}$. Then the complex baseband received signal at the receiver is 
\begin{align} \label{eq:system_model}
	{\bf \bar y} = {\bf \bar H}{\bf \bar x} + {\bf \bar v},
\end{align}
where ${\bf \bar v}$ is a complex Gaussian noise vector with zero mean and covariance matrix ${\sigma}^2\mathbf{I}$. We assume a rich-scattering and frequency-flat channel model, in which all elements of ${\bf \bar H}$ are chosen from complex Gaussian random variables with zero mean and unit variance. The transmit vector $\bf \bar x$ is normalized to satisfy the power constraint $E_{\rm s}$, i.e.,
\begin{align}
	{\rm Tr}\left(\mathbb{E}\left[{\bf \bar x}{\bf \bar x}^{H}\right]\right)=E_{\rm s}.
\end{align}
Without loss of generality, the complex-baseband input-output relationship in \eqref{eq:system_model} can be rewritten in real-vector representation as
 \begin{align}
 {\bf y}={\bf H}{\bf x}+{\bf v},
 \end{align}
 where 
  \begin{align}
 {\bf x} \!=\!\left[\!\! \begin{array}{c} {\rm Re}\{\bf{\bar x}\} \\ {\rm Im}\{\bf{\bar x}\} \end{array}\!\! \right]\in \mathbb{R}^{2N_{\rm t}}, {\bf y} \!=\!\left[\!\! \begin{array}{c} {\rm Re}\{\bf{\bar y}\} \\ {\rm Im}\{\bf{\bar y}\} \end{array} \!\!\right]\in \mathbb{R}^{2N_{\rm r} }, {\bf v} \!=\left[\!\! \begin{array}{c} {\rm Re}\{\bf{\bar v}\} \\ {\rm Im}\{\bf{\bar v}\} \end{array} \!\!\right]\in \mathbb{R}^{2N_{\rm r} }, \nonumber
 \end{align} and
 \begin{align}
 {\bf H} =\left[ \begin{array}{cc} {\rm Re}\{\bf{\bar H}\}& -{\rm Im}\{\bf{\bar H}\} \\ {\rm Im}\{\bf{\bar H}\} & {\rm Re}\{\bf{\bar H}\} \end{array} \right ] \in \mathbb{R}^{2N_{\rm r} \times 2N_{\rm t}}. \nonumber
 \end{align}
For notational convenience, we will use this real-value representation of the MIMO system in this paper, except in Section IV. We also assume that $M$-ary pulse-amplitude-modulation (PAM), $\left\{\frac{-M}{2},\frac{-M}{2}+1,\ldots,\frac{M}{2}-1,\frac{M}{2}\right\}$, is used for the input constellation points per dimension; therefore the input signal set is finite. We also assume that perfect channel state information is known to the receiver, i.e., perfect CSIR, which can be reliably estimated using conventional pilot-transmission.

\subsection{Preliminaries}
We provide some useful definitions which also can be found in \cite{ConwayBook, Forney1998}.

{\bf Definition 1 (Lattice):} Let $\left\{{\bf g}_1,\ldots, {\bf g}_n\right\}$ be a set of linearly independent vectors in $\mathbb{R}^n$, in which each vector constitutes a basis for the lattice.  A real lattice is the countably infinite set defined by integer combinations of basis vectors, i.e.,
\begin{align}
	{\bf \Lambda} =\left\{{\bf x}\in \mathbb{R}^n: {\bf x}=c_1{\bf g}_1 +c_2{\bf g}_2+\cdots +c_n{\bf g}_n\right\},
\end{align}
where $c_i\in \mathbb{Z}$. Thus, a lattice, ${\bf \Lambda}$, is a discrete additive subgroup of $\mathbb{R}^n$, and is closed under addition and reflection. Matrix ${\bf G}=\left[{\bf g}_1,\ldots, {\bf g}_n\right]^{\top}$ is referred to as a generating matrix of the lattice.

{\bf Definition 2 (Nearest neighbor quantizer):} The nearest-neighbor quantizer associated with ${\bf \Lambda}$ is defined as
\begin{align}
	Q({\bf v})={\bf x}_i \in  {\bf \Lambda}~ {\rm if}~ {\bf v}\in\left\{{\bf v}\in \mathbb{R}^n: \|{\bf v}-{\bf x}_i \|_2 \leq \|{\bf v}-{\bf x}_j\|_2\right\},
\end{align}
for any other point ${\bf x}_j\in  {\bf \Lambda}$. The Voronoi cell associated with ${\bf x}_i \in {\bf \Lambda}$ is the set of points in $\mathbb{R}^n$ closest to ${\bf x}_i$, i.e.,$\mathcal{V}_i({\bf \Lambda})=\{{\bf v}: Q({\bf v}) ={\bf x}_i\}$.

{\bf Definition 3 (Nested Lattice):} A  pair of $n$-dimensional  lattices $({\bf \Lambda}_1,{\bf \Lambda}_2) $ is  called  nested  if ${\bf \Lambda}_2 \subset {\bf \Lambda}_1$,  i.e., corresponding generator matrices ${\bf G}_1$ and ${\bf G}_2$ exist such that ${\bf G}_2={\bf G}_1{\bf A}$, where ${\bf A}$ is an $n\times n$ integer matrix that has a determinant $>1$. The volumes of the Voronoi cells of ${\bf \Lambda}_1$ and ${\bf \Lambda}_2$ satisfy $\mathcal{V}_i({\bf \Lambda}_1) = \det({\bf A})\mathcal{V}_i({\bf \Lambda}_2)$.

{\bf Definition 4 (Normalized Coding Gain):} Let $d_{\rm min}^2({\bf \Lambda})$ be the minimum squared distance of the lattice points in ${\bf \Lambda}$. The Hermite parameter of ${\bf \Lambda}$ is the normalized density parameter or the normalized coding gain, which is defined as
\begin{align}
	\gamma_c({\bf \Lambda}) = \frac{d_{\rm min}^2({\bf \Lambda})}{ \mathcal{V}_0({\bf \Lambda})^{\frac{2}{n}}}.
\end{align}
Asymptotically, for very high SNR, determining the maximum possible nominal coding gain of an $n$-dimensional lattice code is equivalent to finding the densest lattice in a sphere-packing sense.

\section{Spatial Lattice Modulation}
In this section, we present the idea of SLM. Unlike the existing SM techniques in which information bits are separately mapped to a set of antenna indices and modulation symbols, the key idea of SLM is to jointly map $K$ information bits to one of $2^K$ lattice vectors in $\mathbb{R}^{2N_{\rm t}}$. This joint mapping strategy using lattices makes it possible to obtain the maximum entropy of input symbol vectors for an given $M$-ary PAM condition per dimension. Also, by using a dense lattice, we are able to achieve the largest nominal coding gain for a given $N_{\rm t}$. Depending on different lattice structures, we propose two SLM methods: SLM using a cubic lattice and SLM using a dense lattice.
\subsection{SLM using Cubic Lattices}
 This proposed SLM method uses a joint mapping strategy to map $K$ information bits into a set of information symbol vectors, each in a cubic lattice. Let $\mathcal{S}^{\rm CB}=\left\{{\bf s}_1,{\bf s}_2,\ldots, {\bf s}_{2^K}\right\}$ be a set of transmit symbol vectors where ${\bf s}_k \in \mathbb{R}^{2N_{\rm t}}$ is the real-representation of ${\bf \bar s}_k\in\mathbb{C}^{N_{\rm t}}$. We denote the number of activated dimensions in an $2N_{\rm t}$-dimensional space by $N_{\rm a}$, where $0\leq N_{{\rm a}}\leq 2N_{\rm t}$. Under the premise that $N_{\rm a}$ dimensions are active, a set of all possible transmit vectors using generalized spatial modulation method under the constraint of the $M$-ary PAM input signal per dimension is 
\begin{align}
	\mathcal{S}^{N_{\rm a}}=\left\{{\bf s}_{1}^{N_{{\rm a}}}, \ldots, {\bf s}_{L_{N_{\rm a}}}^{N_{{\rm a}}} \right\},
\end{align}
where the cardinality of $\mathcal{S}^{N_{\rm a}}$ is
\begin{align}
	L_{N_{\rm a}} ={2N_{\rm t} \choose  N_{\rm a}}M^{N_{\rm a}}.
\end{align}
Because $0\leq N_{{\rm a}}\leq 2N_{\rm t}$, we construct an entire signal set for the joint mapping by the union of $\mathcal{S}^{N_{\rm a}}$ as
\begin{align}
	\mathcal{S}^{\rm CB}(N_{\rm t},M)=\cup_{N_{\rm a}=0}^{2N_{\rm t}}\mathcal{S}^{N_{\rm a}}.
\end{align} 
$\mathcal{S}^{i}$ and $\mathcal{S}^{j}$ are disjoint for all $i\neq j$, so the cardinality of $\mathcal{S}^{\rm CB }$ is
 \begin{align} \label{eq:SLM_num}
 |\mathcal{S}^{\rm CB}(N_{\rm t},M)|=\sum_{i=0}^{2N_{\rm t}}{2N_{\rm t} \choose i}M^i  = (M+1)^{2N_{\rm t}},
 \end{align}
 where the last equality follows from the binomial expansion.
 
 \vspace{0.1cm}
{\bf Example 1:} Suppose $N_{\rm t}=2$ and $M=2$. In this case, we can create 81 lattice vectors by using the joint mapping strategy. The corresponding 81 vectors are listed in Table I\footnote{In case of high bit-rates transmission, look-up table method is not suitable to implement practical system. Alternatively, we can avoid this look-up table by using combinatorial method \cite{Buckles1977}.}. These cubic lattice vectors can be also generated by using the following generating matrix of ${\bf \Lambda}^{{\rm CB}}_4$, i.e., 
\begin{align}
	{\bf G}^{{\rm CB}}_4= \left[ \begin{array}{cccc}
   1 & 0 &  0 & 0 \\
   0 & 1 &  0 & 0 \\
   0 & 0 &  1 & 0 \\
   0 & 0 &  0 & 1 \\
   \end{array}  \right].
\end{align}
Using this set of vectors, one can send information bits at $\log_2(81)=6.3$ bits/sec/Hz if SNR is high enough.
\begin{table}
\centering
	\begingroup
\caption{Look-up Table for SLM with Cubic lattices $\left( N_{\rm t} = 2, ~M = 2 \right)$.}
\begin{tabular}{||c|c|c||c|c|c||}
\Xhline{3\arrayrulewidth}
Index & $\bf x^{\rm T}$ &  $\mathcal{S}^{N_{\rm a}}$  & Index & $\bf x^{\rm T}$ & $\mathcal{S}^{N_{\rm a}}$\\
\Xhline{3\arrayrulewidth}
1& ${[ 0,   0,   0,   0 ]}$ & $\mathcal{S}^{0}$ & 22-25 & ${[ 0,   \pm 1,   \pm 1,   0 ]}$ & \multirow{3}{*}{$\mathcal{S}^{2}$} \\ \cline{1-3} \cline{4-5}
2-3 & ${[ \pm 1,   0,   0,   0 ]}$ & \multirow{4}{*}{$\mathcal{S}^{1}$} & 26-29 & ${[ 0,   \pm 1, 0  ,   \pm 1 ]}$ & \\ \cline{1-2} \cline{4-5}
4-5 & ${[ 0,   \pm 1,   0,   0 ]}$ &  & 30-33 & ${[ 0,   0, \pm 1  ,   \pm 1 ]}$ &  \\ \cline{1-2} \cline{4-6}
6-7 & ${[ 0,   0,   \pm 1,   0 ]}$ &  & 34-41 & ${[ \pm 1,   \pm 1, \pm 1  ,  0 ]}$ & \multirow{4}{*}{$\mathcal{S}^{3}$} \\ \cline{1-2} \cline{4-5}
8-9 & ${[ 0,   0,   0,   \pm 1 ]}$ &  & 42-49 & ${[ \pm 1, \pm 1, 0 ,  \pm 1 ]}$ & \\ \cline{1-3} \cline{4-5}
10-13& ${[ \pm 1,   \pm 1,   0,   0 ]}$ & \multirow{3}{*}{$\mathcal{S}^{2}$} & 50-57 & ${[ \pm 1,   0,   \pm 1,   \pm 1 ]}$ &\\ \cline{1-2} \cline{4-5}
14-17 & ${[ \pm 1,   0,   \pm 1,   0 ]}$ &  & 58-65 & ${[ 0,   \pm 1,   \pm 1,   \pm 1 ]}$ &  \\ \cline{1-2} \cline{4-6}
18-21 & ${[ \pm 1,   0,   0,   \pm 1 ]}$ &  & 66-81 & ${[ \pm 1,   \pm 1,   \pm 1,   \pm 1 ]}$ & $\mathcal{S}^{4}$ \\ \cline{1-2} 
\hline
\Xhline{3\arrayrulewidth}
\end{tabular} \label{table:SLM_CB}
\endgroup
\end{table}

\vspace{0.1cm}
{\bf Proposition 1:} Let ${\bf x}\in \mathbb{R}^{2N_{\rm t}}$ be an input symbol vector that is chosen uniformly from the proposed signal set $\mathcal{S}^{\rm CB}(N_{\rm t},M)$. Then the maximum input entropy of the proposed signal set is 
\begin{align}
	\mathcal{H}({\bf x}) = 2N_{\rm t}\log_2 (M+1).
\end{align} 

\begin{proof}
	The proof follows from the definition of the entropy and the uniform input distribution, i.e., $P({\bf x}={\bf s}_k)=\frac{1}{(M+1)^{2N_{\rm t}}}$ for $k\in\left\{1,2\ldots, (M+1)^{2N_{\rm t}}\right\}$.
\end{proof}
We offer some remarks and examples on our joint mapping strategy to clarify understanding.

\vspace{0.1cm}
 {\bf Remark 1:} The proposed joint mapping method has an effective modulation size of $M+1$ per dimension. Because we assume that $M$-ary PAM is used for the input constellation set per dimension, the modulation size of $M$ is trivial. However, because we use spatial modulation, we can add one hidden constellation point of `0` as an input vector by physically deactivating the input per dimension; this process provides the effective modulation size of $M+1$. This increment provides a considerable gain in the input entropy. For example, suppose a spatial multiplexing transmission method with uniform $M$-ary PAM input signaling. The maximum input entropy of this method is $2N_{\rm t}\log_2(M)$. Therefore, one can achieve the gain of
 \begin{align}
 	2N_{\rm t}\log_2(M+1)-2N_{\rm t}\log_2(M)=2N_{\rm t}\log_2\left(1+\frac{1}{M}\right). \nonumber
 \end{align}
 This entropy gain implies that one can send $2N_{\rm t}\log_2\left(1+\frac{1}{M}\right)$ more information bits per transmission using the proposed method when SNR is high enough. The gain increases linearly with $N_{\rm t}$ and is therefore unbounded.
 
\vspace{0.1cm}
 {\bf Remark 2:} The proposed joint mapping technique generalizes the existing SM methods in \cite{Mesleh2008,Jeganathan2009,Younis2010,Wang2012,Mesleh2015,Guo2016} and spatial multiplexing method. For example, supposing that $N_{\rm t}=2$ and $M=2$, the signal sets generated by the conventional SM and QSM method in \cite{Mesleh2008,Mesleh2015} are contained in $\mathcal{S}^2$, which is a subset of $\mathcal{S}^{\rm CB}(2,2)$. Similarly, the signal set of the conventional spatial multiplexing method is $\left\{1+j,1-j,-1+j,-1-j\right\}^2$, which is same with $\mathcal{S}^4$ of $\mathcal{S}^{\rm CB}(2,2)$.

\vspace{0.1cm}
  {\bf Remark 3:} When the effective modulation size $M+1$ is a prime number, the proposed constellation set $\mathcal{S}^{\rm CB}(N_{\rm t},M)$ generated by the joint spatial mapping can also be constructed by a nested lattice modulation method. Letting $\epsilon = \frac{1}{\sqrt{E_s}}$, we consider two nested $2N_{\rm t}$-dimensional cubic lattices with the generating matrices ${\bf G}_c=\epsilon{\bf I}_{2N_{\rm t}\times 2N_{\rm t}}$ and ${\bf G}_s=\epsilon (M+1) {\bf I}_{2N_{\rm t} \times 2N_{\rm t}}$ respectively:,
  \begin{align}
  {\bf 	\Lambda}_c &= \left\{{\bf x}\in \mathbb{R}^{2N_{\rm t}}: {\bf x}={\bf c}^{\top} {\bf G}_c\right\}~~{\rm and} \nonumber \\
  	{\bf 	\Lambda}_s&= \left\{{\bf x}\in \mathbb{R}^{2N_{\rm t}}: {\bf x}={\bf c}^{\top} {\bf G}_s\right\},
  \end{align}
where ${\bf c}\in\mathbb{Z}^{2N_{\rm t}}$ and ${\bf 	\Lambda}_c \subset {\bf 	\Lambda}_s$. Using these nested lattices, the same constellation set is generated by \begin{align}
	\mathcal{S}^{\rm CB}(N_{\rm t},M)=  {\bf \Lambda}_c\cap \mathcal{V}_0({\bf \Lambda}_s),
\end{align}
where $\mathcal{V}_0({\bf \Lambda}_s)$ is the Voronoi region associated with ${\bf 0}\in {\bf \Lambda}_s$, i.e., the volume $\left[-\frac{\epsilon (M+1)}{2}, \frac{\epsilon (M+1)}{2}\right)^{2N_{\rm t}}$. Therefore, the joint spatial mapping method can be interpreted as the nested lattice modulation using the corresponding cubic generating matrices.
\subsection{SLM with Low-Dimensional Dense Lattices}
The cubic lattice ${\bf \Lambda}^{\rm CB}$ used for the previous SLM is a baseline lattice, because by the definition it offers no nominal coding gain, i.e., $\gamma_c\left({\bf \Lambda}^{\rm CB}\right)=1$. Therefore, a natural extension is to use a dense lattice ${\bf \Lambda}$ to construct a set of multi-dimensional lattice vectors that yield a higher nominal coding gain than ${\bf \Lambda}^{\rm CB}$ in a given number of dimensions, i.e., $\gamma_c\left({\bf \Lambda}\right)>\gamma_c\left({\bf \Lambda}^{\rm CB}\right)$.

Finding the densest lattice packing in an arbitrary number of dimensions is a difficult mathematical problem. The lattices that have the largest nominal coding gain are well characterized up to 24 dimensions \cite{ConwayBook,Forney1998,Tarokh1999}. The Barnes-Wall lattice is a good one due to its simplicity of lattice construction and its tractability to analyze. It also provides high nominal coding gains in a low-dimensional signal space.

The following lemma provides a method to construct the Barnes-Wall lattice in ${\mathbb{R}}^{2^{m+1}}$\cite{Nebe2002}; this lattice is used in the proposed SLM. 

\vspace{0.1cm}
\begin{lem}[Barnes-Wall Lattice Construction \cite{Nebe2002}] Let ${\bf M}_1$ denote the generating matrix of a balanced Barnes-Wall lattice in $\mathbb{R}^2$:
\begin{align}
 	{\bf M}_1=\left[ \begin{array}{cc} \sqrt2& 0 \\ 1 & 1 \end{array} \right ].
 \end{align}
The generating matrix of ${\bf \Lambda}^{{\rm BW}}_{2^{m+1}}$ lattice in ${\mathbb{R}}^{2^{m+1}}$ is obtained from ${\bf M}_{m+1}$, which is the $m+1$ times Kronecker products of ${\bf M}_1$, i.e., 
\begin{align}
 	{\bf M}_{m+1}=\underbrace{{\bf M}_1\otimes{\bf M}_1\otimes\cdots\otimes {\bf M}_1}_{m+1}.
 \end{align}
By rescaling the irrational elements in ${\bf M}_{m+1}$,  the generating matrix of ${\bf \Lambda}^{{\rm BW}}_{2^{m+1}}$ is obtained as
 \begin{align}
 	{\bf G}^{\rm BW}_{2^{m+1}}(i,j)&=\begin{cases}
        {\bf M}_{m+1}(i,j), & {\rm if}~ {\bf M}_{m+1}(i,j) ~\text{is rational}, \\
        \frac{1}{\sqrt2}{\bf M}_{m+1}(i,j), &{\rm if}~ {\bf M}_{m+1}(i,j)~\text{is irrational}.\label{eq:BW_gen}
    \end{cases}
 \end{align}
 \end{lem}
The following lemma shows some coding theoretic properties of Barnes-Wall lattices.
  
\vspace{0.1cm}
\begin{lem}[Coding Properties of Barnes-Wall Lattices \cite{ConwayBook,Forney1998}]
		For all integer $m \geq 0$, a $2^{m+1}$-dimensional Barnes-Wall lattice ${\bf \Lambda}^{{\rm BW}}_{2^{m+1}}$ exists that has the minimum squared Euclidean distance $d^2_{\rm min}\left({\bf \Lambda}^{{\rm BW}}_{2^{m+1}}\right) = 2^m$ and the normalized volume $\mathcal{V}\left({\bf \Lambda}^{{\rm BW}}_{2^{m+1}}\right)^\frac{1}{2^m}=2^{\frac{m}{2}}$; therefore its nominal coding gain is $\gamma_c\left({\bf \Lambda}^{{\rm BW}}_{2^{m+1}}\right)=2^{\frac{m}{2}}$. In addition, the kissing number of ${\bf \Lambda}^{{\rm BW}}_{2^{m+1}}$ is $K_{\rm min}\left({\bf \Lambda}^{{\rm BW}}_{2^{m+1}}\right) = \prod_{i=1}^{m+1}\left(2^i+2\right)$.		
\end{lem}

\begin{algorithm}[t]\small
    \caption{Signal Set Design Method for SLM using Dense Lattices.}
  \begin{algorithmic}[1]
    \INPUT Generating matrix ${\bf G} \in \mathbb{Z}^{2N_{\rm t} \times 2N_{\rm t}} $, Maximum power constraint $P_{\rm max}$.
    \OUTPUT Signal set for SLM $\mathcal{S}^{\rm BW}(N_{\rm t},P_{\rm max})$ $\subset$  $\mathbb{Z}^{2N_{\rm t}}$.
    \STATE \textbf{Initialization} $P=0$.
    \FOR {$P \in \left[0,1,2,\cdots, P_{\rm max} \right]$}
      \STATE {Define $2N_{\rm t}$ dimensional integer row vector ${\bf{s}} = [s_1, s_2, \ldots, s_{2N_{\rm t}}]$. Then, exhaustively search ${\bf{s}}$ satisfying the power condition $P$.}
      \STATEx  $\tilde{\mathcal{S}}^{P} := \{  {\bf{s}} ~| ~{ \| \bf{s} \|^2_2} = P ,~ {\bf s} \in \mathbb{Z}^{2N_{\rm t}}~ ,~ |{s}_i| \geq |{s}_j| ~ ,\forall~ i \geq j \}$.
     \STATE Let $\hat{\mathcal{S}}^{P}$ denote the symmetric group, or group of permutations, on $\tilde{{\mathcal{S}}}_P$.
     \STATEx $\hat{\mathcal{S}}^{P} := \{Sym \{ [{{s}}_1, {{s}}_2, \ldots, {{s}}_{2N_{\rm t}}] \}, ~\forall {\bf{s}} \in \tilde{\mathcal{S}}^{P}\}$. 
     \STATE {Check whether ${\hat{\bf{s}}}$ is in lattice $\Lambda$.}
     \STATEx $\mathcal{S}^{P} :=\{ {\hat{\bf{s}}} ~| ~{\hat{\bf{s}}}\in \hat{\mathcal{S}}_P , ~{\hat{\bf{s}}} {\bf{G}}^{-1} \in \mathbb{Z}^{2N_{\rm t}} \}.$.
    \ENDFOR
    \STATEx Merge all $\mathcal{S}^{P}$ sets into $\mathcal{S}^{\rm BW}$.
    \STATE $\mathcal{S}^{\rm BW}(N_{\rm t},P_{\rm max}) = \cup_{P=0}^{P_{\rm max}}\mathcal{S}^{P}.$
  \end{algorithmic}
\end{algorithm}

By exploiting the Barnes-Wall lattice ${\bf \Lambda}^{{\rm BW}}_{2^{m+1}}$ that is constructed by ${\bf G}^{\rm BW}_{m+1}$ in \eqref{eq:BW_gen}, we propose Algorithm 1, which creates a signal set for SLM using dense lattices. The algorithm finds a signal set $\mathcal{S}^{\rm BW}$ such that each lattice vector is created by generating matrix ${\bf G} \in \mathbb{Z}^{2N_{\rm t} \times 2N_{\rm t}} $ and satisfies the maximum power constraint $P_{\rm max}$. The first step is to exhaustively search all possible integer vectors that satisfy the given power $P$. For a vector satisfying the power constraint, the symmetric group is selected as a candidate. For example, with ${\bf \Lambda}^{{\rm BW}}_4$ lattice at power condition $P=2$, $[$$\pm$1,$\pm$1,0,0$]$, $[$$\pm$1,0,$\pm$1,0$]$, $[$ $\pm1,$0,0,$\pm1$ $]$, $[$0,$\pm$1,$\pm$1,0$]$, $[$0,$\pm$1,0,$\pm$1$]$, $[$0,0,$\pm1$,$\pm$1$]$ will be the possible candidates for constellation vectors. After collecting the candidates, we check the condition whether they consist of an integer linear combination of the basis $\left\{{\bf g}_1,\ldots, {\bf g}_{2N_{t}}\right\}$. Then, vectors that satisfy this condition are included in our SLM signal set. Algorithm 1 iterates this procedure by increasing $P$ until the power condition $P$ reaches the maximum power constraint $P_{\rm max}$.

\vspace{0.1cm}
{\bf Example 2:} Suppose a four-dimensional Barnes-Wall lattice, ${\bf \Lambda}^{{\rm BW}}_4$, which is equivalent to the $D_4$ lattice. The generating matrix of ${\bf \Lambda}^{{\rm BW}}_4$ is
\begin{align}
	{\bf G}^{{\rm BW}}_4= \left[ \begin{array}{cccc}
   2 & 0 &  0 & 0 \\
   1 & 1 &  0 & 0 \\
   1 & 0 &  1 & 0 \\
   1 & 1 &  1 & 1 \\
   \end{array}  \right].
\end{align}
The minimum squared Euclidean distance $d^2_{\rm min}\left({\bf \Lambda}^{{\rm BW}}_4\right) = 2$, the volume $\mathcal{V}\left({\bf \Lambda}^{{\rm BW}}_4\right)=|\det\left({\bf G}^{{\rm BW}}_4\right)|=2$, and therefore its nominal coding gain is $\gamma_c\left({\bf \Lambda}^{{\rm BW}}_4\right)=\sqrt{2} \simeq 1.5$ dB. Using this generating matrix and Algorithm 1, we can generate 145 transmit signal vectors in ${\bf \Lambda}^{{\rm BW}}_4$ with $P_{\rm max} = 6 $. The corresponding 145 vectors are listed in Table II\footnote{ The barnes-Wall lattice vectors of Table II can be indexed by using arithmetic operators without look-up table \cite{Rault2001}. This enables to reduce encoding and decoding complexity.}.

\begin{table}[t!]
\centering
	\begingroup
	\caption{Look-up Table for SLM with Barnes-Wall lattices $\left( N_{\rm t} = 2, ~ P_{\rm max} = 6 \right)$.}
\begin{tabular}{||c|c|c||c|c|c||}
\Xhline{3\arrayrulewidth}
Index & $\bf x^{\rm T}$ &  $\mathcal{S}^{P}$  & Index & $\bf x^{\rm T}$ & $\mathcal{S}^{P}$\\
\Xhline{3\arrayrulewidth}
1& ${[ 0,   0,   0,   0 ]}$ & $\mathcal{S}^{0}$ & 46-47 & ${[ 0,   0  , \pm 2,   0 ]}$ & \multirow{2}{*}{$\mathcal{S}^{4}$} \\ \cline{1-3} \cline{4-5}
2-5 & ${[ \pm 1,   \pm 1,   0,   0 ]}$ & \multirow{6}{*}{$\mathcal{S}^{2}$} & 48-49 & ${[ 0,   0, 0  ,   \pm 2]}$ & \\ \cline{1-2} \cline{4-6}
6-9 & ${[ \pm 1,   0,   \pm 1,   0 ]}$ &  & 50-57 & ${[ \pm 2,   \pm 1, \pm 1  ,  0 ]}$ & \multirow{8}{*}{$\mathcal{S}^{6}$} \\ \cline{1-2} \cline{4-5}
10-13 & ${[ \pm 1,   0,   0,   \pm 1 ]}$ &  & 58-65 & ${[ \pm 1,   \pm 2, \pm 1  ,  0 ]}$ &  \\ \cline{1-2} \cline{4-5}
14-17 & ${[ 0,   \pm 1,   \pm 1,   0 ]}$ &  & 66-73 & ${[ \pm 1, \pm 1, \pm 2 ,  0]}$ & \\ \cline{1-2} \cline{4-5}
18-21& ${[ 0,   \pm 1,   0,   \pm 1 ]}$ &  & \multirow{2}{*}{\vdots} & \multirow{2}{*}{$\vdots$} &\\ \cline{1-2}
22-25 & ${[ 0 ,   0,   \pm 1,   \pm 1 ]}$ &  &  &  &  \\ \cline{1-3} \cline{4-5}
26-41 & ${[ \pm 1,   \pm 1,   \pm 1,   \pm 1 ]}$ & \multirow{3}{*}{$\mathcal{S}^{4}$} & 122-129 & ${[ 0,   \pm 2,   \pm 1,   \pm 1]}$ &  \\ \cline{1-2} \cline{4-5}
42-43 & ${[ \pm 2,   0,   0,   0 ]}$ &  & 130-137 & ${[ 0,   \pm 1,   \pm 2,   \pm 1]}$ & \\ \cline{1-2} \cline{4-5}
44-45 & ${[0,    \pm 2,   0,   0 ]}$ &  & 138-145 & ${[ 0,   \pm 1,   \pm 1,   \pm 2]}$ &  \\ \cline{1-2} \cline{4-5}
\hline
\Xhline{3\arrayrulewidth}
\end{tabular} \label{table:SLM_CB}
\endgroup
\end{table}

\vspace{0.1cm}
{\bf Remark 4:} The proposed algorithm is valid when the generating matrix has integer elements. Therefore, except $A_2 \subset \mathbb{R}^2$, we are able to construct a signal set for the SLM with any low-dimensional dense lattices as ${\rm BW}_{2^{m+1}}$ and Leech lattice $L_{24}$. As an extreme case, when the channel matrix ${\bf H}\in \mathbb{R}^{2N_{\rm r}\times 2N_{\rm t}}$ is chosen from the set of unitary matrices, the maximum nominal coding gains $\gamma_c({\bf \Lambda})$ achieved by the known lattices are known (Table III) \cite{Tarokh1999}. For a general MIMO channel, however, these gains are not obtainable because the minimum distance of the signal set seen by the receiver diminishes when the channel matrix is non-unitary.

\begin{table}
\centering
	\begingroup
	\caption{The Maximum Nominal Coding Gains for SLM with Low-dimensional Dense Lattices.}
\begin{tabular}{|c|c|c|c|}
\Xhline{3\arrayrulewidth}
Antennas & Dimensions &  Lattice $\Lambda$  & Coding gain $\gamma_c(\Lambda)$ \\
\Xhline{3\arrayrulewidth}
1& 2 & $A_2$ & 0.62 dB \\ \hline
2 & 4 & ${\bf \Lambda}_4^{\rm BW}(=D_4)$ & 1.51 dB \\ \hline
4 &8 & ${\bf \Lambda}_8^{\rm BW}(=E_8)$ & 3.01 dB \\ \hline
8& 16 & ${\bf \Lambda}_{16}^{\rm BW}(=\Lambda_{16})$ & 4.52 dB \\ \hline
12& 24 & $L_{24}$ & 6.02 dB \\ \hline
16& 32 & ${\bf \Lambda}_{32}^{\rm BW}(=\Lambda_{32})$ & 6.28 dB \\ \hline
\Xhline{3\arrayrulewidth}
\end{tabular} \label{table:coding_gain}
\endgroup
\end{table}

\section{Performance Analysis}
This section analyzes the average mutual information and average symbol-vector-error-probability (ASVEP) for the MIMO systems when input symbol vectors are uniformly drawn from an arbitrary finite input constellation $\mathcal{S}$. For ease of exposition, we use the complex baseband MIMO model in \eqref{eq:system_model}.

\subsection{Maximum Likelihood (ML) Detection}
We assume that the receiver and transmitter share the joint mapping rule between information bits and the SLM signal set. The active dimension and constellation points of a transmit symbol vector are jointly decoded using ML principle
\begin{align} \label{eq:ML}
[l]=\arg\max_{m}f_{\bf \bar y}({\bf \bar y}|{\bf \bar x}_m)=\arg\min_{m}\|{\bf \bar y}-{\bf \bar H}{\bf \bar x}_m\|^2,
\end{align}
for ${\bf \bar x}_m\in\mathcal{\bar S}=\{{\bf \bar s}_1, {\bf \bar s}_2, \ldots, {\bf \bar s}_{|\mathcal{\bar{S}}|}\}$. 
The conditional probability density function
 (PDF) of ${\bf \bar y}$ given that ${\bf \bar x}_m$ is  
\begin{align}
f({\bf \bar y}|{{\bf \bar x}_m})=\frac{1}{{({\pi}{{\sigma}^2})}^{N_{\rm r}}}\exp\left(-\frac{{\|{\bf \bar y}-{\bf \bar H}{{\bf \bar x}_m}\|^2}}{{{\sigma}^2}}\right).
\end{align}
We use the ML criterion for the performance analysis and for the comparison with other SM schemes.

\subsection{Average Mutual Information Analysis for SLM}

When input signal vectors are uniformly drawn from a finite constellation set (e.g., $M$-ary PAM, $M$-ary QAM), an exact expression of the mutual information with perfect CSIR was derived in \cite{Guo2014} when the channel is fixed. For the completeness of our analysis, we first present the exact expression of the mutual information. Then, to improve intuition, we also provide a tight approximation expression for the average mutual information.

\subsubsection{Exact Expression}
For a MIMO channel, the mutual information between the discrete input vector ${\bf \bar x}$ and the continuous channel output vector ${\bf \bar y}$ is
\begin{align}\label{eq:MI_Def}
    \mathcal{I}({\bf \bar x};{\bf \bar y}|{\bf \bar H})    &=\mathcal{H}({\bf \bar y}|{\bf \bar H})-\mathcal{H}({\bf \bar y}|{\bf \bar x},{\bf \bar H}) \nonumber\\
    &=\mathcal{H}({\bf \bar y}|{\bf \bar H})- \mathcal{{H}({\bf \bar v})}    \nonumber\\
    &=\mathcal{H}({\bf \bar y}|{\bf \bar H})- N_{\rm r}\log_2({\pi}e{{\sigma}^2}).
\end{align}
Because the received signal vector ${\bf \bar y}$ follows a Gaussian distribution for given ${\bf \bar H}$ and ${\bf \bar x}$, the PDF of ${\bf \bar y}$ is 
\begin{align}\label{eq:distribution_y}
f({\bf \bar y}|{\bf \bar H})=\frac{1}{|\mathcal{S}|}\sum_{{\bf \bar x}_{i}\in  \mathcal{S}}\frac{1}{{({\pi}{{\sigma}^2})}^{N_{\rm r}}}\exp\left(-\frac{{\|{\bf \bar y}-{\bf \bar H}{{\bf \bar x}_i}\|^2}}{{{\sigma}^2}}\right).
\end{align}
Using $f({\bf \bar y}|{{\bf \bar x}_i},{\bf \bar H})$ and $f({\bf \bar y}|{\bf \bar H})$, the mutual information for given $\bf \bar H$ is  
\begin{align}\label{eq:mutual information x,y}
\mathcal{I}({\bf \bar x};{\bf \bar y}|{\bf \bar H})=\sum_{{\bf \bar x}_{i}\in  \mathcal{S}}\frac{1}{|\mathcal{S}|}\int_{{\bf \bar y}}f({\bf \bar y}|{\bf \bar x}={\bf \bar x}_i,{\bf \bar H})\log_2\frac{f({\bf \bar y}|{\bf \bar x}={\bf \bar x}_i,{\bf \bar H})}{f({\bf \bar y}|{\bf \bar H})}d{\bf \bar y}.
\end{align}
Using equations (21)-(23), the mutual information for SLM is obtained as \begin{align}\label{eq:MI}
&\mathcal{I}({\bf \bar x};{\bf \bar y}|{\bf \bar H}) =\log{|\mathcal{S}|}  \nonumber \\
&-\frac{1}{{|\mathcal{S}|}}\!\sum_{{\bf \bar x}_{i}\in  \mathcal{S}}{{\mathbb{E}}_{\bf \bar v}}\left[\log_2\sum_{{\bf \bar x}_{j}\in  \mathcal{S}}\exp\left(-\frac{\|{\bf \bar H}({\bf \bar x}_{i}-{\bf \bar x}_{j})\!+\!{\bf \bar v}\|^2\!-\!\|{\bf \bar v}\|^2}{{{\sigma}^2}}\!\right) \right].
\end{align}

\subsubsection{Approximation of Average Mutual Information}
The derived mutual information expression in \eqref{eq:MI} is not tractable because it involves the multi-dimensional integral with respect to ${\bf \bar v}$ and it does not capture the randomness in channels. We derive a closed form tight approximation of the average mutual information by exploiting the lower bound of \eqref{eq:MI}. The following proposition closely approximates the average mutual information. 

\vspace{0.1cm}
{\bf Proposition 3:} The average mutual information, $\mathbb{E}_{\bf \bar H}\left[\mathcal{I}\left(\bf \bar x ; \bf \bar y| \bf \bar H \right)\right]$ for SLM, can be approximated as
\begin{align}\label{eq:approx}
\mathbb{E}_{\bf \bar H}&\left[\mathcal{I}\left(\bf \bar x ; \bf \bar y| \bf \bar H \right)\right]\nonumber\\&\approx2\log_2|\mathcal{S}|-\log_2\sum_{{\bf x}_{i}\in  \mathcal{S}}\sum_{{\bf x}_{j}\in  \mathcal{S}}\left(1+\frac{\|{\bf \bar x}_{i}-{\bf \bar x}_{j}\|^2}{2{\sigma}^2}\right)^{-N_r}.
\end{align}

\begin{proof}
Applying Jensen's inequality yields a lower bound of $\mathcal{H}({\bf \bar y}|{\bf \bar H})$ as in \cite{Guo2014}, 
\begin{align}\label{eq:lowerbound_y}
&\mathcal{H}({\bf \bar y}|{\bf \bar H})\geq-\log_2{\mathbb{E}}_{\bf \bar y}\left[f({\bf \bar y}|{\bf \bar H})\right]= \nonumber \\
&-\log_2\int_{\bf \bar y}\left(\frac{1}{|\mathcal{S}|}\sum_{{\bf \bar x}_{i}\in  \mathcal{S}}\frac{1}{{({\pi}{{\sigma}^2})}^{N_{\rm r}}}\exp\left(-\frac{{\|{\bf \bar y}-{\bf \bar H}{{\bf \bar x}_i}\|^2}}{{{\sigma}^2}}\right)\right)^2d{\bf \bar y}.
\end{align}
By changing the order of summation and integration in \eqref{eq:lowerbound_y}, $\mathcal{H}({\bf \bar y}|{\bf \bar H})$ is lower-bounded by
\begin{align}\label{eq:lowerbound_y2}
\mathcal{H}({\bf \bar y}|{\bf \bar H})&\geq2\log_2|\mathcal{S}|(\pi{\sigma}^2)^{N_{\rm r}}\nonumber\\
&-\log_2\sum_{{\bf \bar x}_{i}\in  \mathcal{S}}\sum_{{\bf \bar x}_{j}\in  \mathcal{S}}\int_{\bf \bar y}{e^{-\frac{{\|{\bf \bar y}-{\bf \bar H}{{\bf \bar x}_{i}}\|^2}}{{{\sigma}^2}}}}{e^{-\frac{{\|{\bf \bar y}-{\bf \bar H}{{\bf \bar x}_{j}}\|^2}}{{{\sigma}^2}}}}d{\bf \bar y}.
\end{align}
The integration in \eqref{eq:lowerbound_y2} can be calculated as
\begin{align}\label{eq:lowerbound_y3}
\int_{\bf \bar y}{e^{-\frac{{\|{\bf \bar y}-{\bf \bar H}{{\bf \bar x}_{i}}\|^2}}{{{\sigma}^2}}}}{e^{-\frac{{\|{\bf \bar y}-{\bf \bar H}{{\bf \bar x}_{j}}\|^2}}{{{\sigma}^2}}}}d{\bf \bar y}=e^{-\frac{\|{\bf \bar H}{{\bf \bar x}_{i}}\|^2+\|{\bf \bar H}{{\bf \bar x}_{j}}\|^2}{{{\sigma}^2}}}\times \nonumber\\
\!\!\!\!\underbrace{\int_{\bf \bar y}\exp\left(-\frac{2\|{\bf \bar y}\|^2-2{\rm Re}\left\{{\bf \bar y}^{H}({\bf \bar H}{\bf \bar x}_{i}+{\bf \bar H}{\bf \bar x}_{j})\right\}}{{{\sigma}^2}}\right)d{\bf \bar y}}_{\Theta}.
\end{align}
Because the elements of ${\bf \bar y}=[y_1, y_2,\ldots, y_{N_{\rm r}}]^{\top}$ are mutually independent, the multi-dimensional integrations in \eqref{eq:lowerbound_y3} are computed using element-wise integrations as
\begin{align}\label{eq:lowerbound_y4}
\Theta=\prod_{k=1}^{N_{\rm r}}\int_{y_k}\exp\left(-\frac{2\|{y_k}\|^2-2{\rm Re}\left\{{y^\ast_k\left\{{\bf \bar H}({\bf \bar x}_{i}+{\bf \bar x}_{j})\right\}_k}\right\}}{{{\sigma}^2}}\right)dy_k.
\end{align}
The integration in \eqref{eq:lowerbound_y4} is conducted separately for real and imaginary parts to yield
\begin{align}\label{eq:lowerbound_y5}
\int_{y_{k,{\rm Re}}}\exp\left(-\frac{2{y^2_{k,{\rm Re}}}-2{{y_{k,{\rm Re}}}\left\{{\bf \bar H}({\bf \bar x}_{i}+{\bf \bar x}_{j})\right\}_{k,{\rm Re}}}}{{{\sigma}^2}}\right)dy_{k,{\rm Re}}  \nonumber\\
\times\int_{y_{k,{\rm Im}}}\exp\left(-\frac{2{y^2_{k,{\rm Im}}}-2{{y_{k,{\rm Im}}}\left\{{\bf \bar H}({\bf \bar x}_{i}+{\bf \bar x}_{j})\right\}_{k,{\rm Im}}}}{{{\sigma}^2}}\right)dy_{k,{\rm Im}}.
\end{align}
The integrals in \eqref{eq:lowerbound_y5} are computed using the following identity
\begin{align}\label{eq:lowerbound_y6}
\int_{-\infty}^{\infty}\exp\left(-\frac{2y^2-2xy}{{\sigma}^2}\right)dy=\sqrt{\frac{\pi\sigma^2}{2}}\exp\left(\frac{x^2}{2\sigma^2}\right).
\end{align}
Thus, $\Theta=\Theta_{\rm Re}\times\Theta_{\rm Im}$, where $\Theta_{\rm Re}$ and $ \Theta_{\rm Im}$ are given by
\begin{align}\label{eq:lowerbound_y7}
\Theta_{\rm Re}=\prod_{k=1}^{N_{\rm r}}\sqrt{\frac{\pi\sigma^2}{2}}\exp\left(\frac{\left\{{\bf \bar H}({\bf \bar x}_{i}+{\bf \bar x}_{j})\right\}^2_{k,\rm Re}}{\sigma^2}\right), \nonumber\\
\Theta_{\rm Im}=\prod_{k=1}^{N_{\rm r}}\sqrt{\frac{\pi\sigma^2}{2}}\exp\left(\frac{\left\{{\bf \bar H}({\bf \bar x}_{i}+{\bf \bar x}_{j})\right\}^2_{k,\rm Im}}{\sigma^2}\right).
\end{align}
Using the results in \eqref{eq:lowerbound_y7}, we obtain the simple 
\begin{align}\label{eq:lowerbound_y8}
\Theta&=\prod_{k=1}^{N_{\rm r}}\left(\frac{\pi{{\sigma}^2}}{2}\right)\exp\left(\frac{\left\|\left\{{\bf \bar H}({\bf \bar x}_{i}+{\bf \bar x}_{j})\right\}_k\right\|^2}{2{{\sigma}^2}}\right)\nonumber \\
&=\left(\frac{\pi{{\sigma}^2}}{2}\right)^{N_{\rm r}}\exp\left({\frac{\|{\bf \bar H}({\bf \bar x}_{i}+{\bf \bar x}_{j})\|^2}{2{{\sigma}^2}}}\right).
\end{align}
Using the above equations \eqref{eq:lowerbound_y2}-\eqref{eq:lowerbound_y8}, the lower bound of $\mathcal{H}({\bf \bar y}|{\bf \bar H})$ is 
\begin{align}\label{eq:lowerbound_y9}
\mathcal{H}({\bf \bar y}|{\bf \bar H})&\geq2\log_2|\mathcal{S}|+\log_2(\pi{\sigma}^2)^{N_{\rm r}}+N_{\rm r}\log_2{2}\nonumber\\
&-\log_2\sum_{{\bf \bar x}_{i}\in  \mathcal{S}} \sum_{{\bf \bar x}_{j}\in  \mathcal{S}} \exp\left(-\frac{{\|{\bf \bar H}({\bf \bar x}_{i}-{\bf \bar x}_{j})\|^2}}{{2{\sigma}^2}}\right).
\end{align}
Plugging \eqref{eq:lowerbound_y9} into \eqref{eq:MI_Def}, a lower bound of $\mathcal{I}({\bf \bar x};{\bf \bar y}|{\bf \bar H})$ is obtained as
\begin{align}\label{eq:lowerbound_y10}
\mathcal{I}^{{\rm Low}}({\bf \bar x};{\bf \bar y}|{\bf \bar H})&=2\log_2{|\mathcal{S}|}+N_{\rm r}\log_2\frac{2}{e} \nonumber \\
&-\log_2\sum_{{\bf \bar x}_{i}\in  \mathcal{S}} \sum_{{\bf \bar x}_{j}\in  \mathcal{S}} \exp\left(-\frac{{\|{\bf \bar H}({\bf \bar x}_{i}-{\bf \bar x}_{j})\|^2}}{{2{\sigma}^2}}\right)\nonumber \\
&=2\log_2{|\mathcal{S}|}+N_{\rm r}\log_2\frac{2}{e} \nonumber \\
&-\log_2\left\{|\mathcal{S}|\!+\!\!\!\sum_{{\bf \bar x}_{i}\in  \mathcal{S}} \sum_{{\bf \bar x}_{j}\in  \mathcal{S}/\{{\bf \bar x}_{i}\}}\!\!\! \!\!\exp\left(-\frac{{\|{\bf \bar H}({\bf \bar x}_{i}-{\bf \bar x}_{j})\|^2}}{{2{\sigma}^2}}\right)\right\}.
\end{align}
From \eqref{eq:lowerbound_y10}, by taking the limits to the two extreme SNR regimes, we obtain the limit values:
\begin{align}\label{eq:limits}
\lim_{{\rm SNR}\rightarrow0}\mathcal{I}^{{\rm Low}}({\bf \bar x};{\bf \bar y}|{\bf \bar H})&=N_{\rm r}\log_2\frac{2}{e},\nonumber \\
\lim_{{\rm SNR}\rightarrow\infty}\mathcal{I}^{{\rm Low}}({\bf \bar x};{\bf \bar y}|{\bf \bar H})&=N_{\rm r}\log_2\frac{2}{e} +\log_2{|\mathcal{S}|}.
\end{align}
Intuitively, the mutual information should approach zero when SNR is very low, but should approach $\log_2{|\mathcal{S}|}$ when SNR is very high. Using these facts and the two extreme values in \eqref{eq:limits}, we define the offset for the approximation as,
\begin{align}
	\triangle_{\mathcal{I}({\bf \bar x};{\bf \bar y}|{\bf \bar H})}=\mathcal{I}^{{\rm Low}}({\bf \bar x};{\bf \bar y}|{\bf \bar H})-\mathcal{I}({\bf \bar x};{\bf \bar y}|{\bf \bar H})=N_{\rm r}\log_2\frac{e}{2}.
\end{align} 
Subtracting this offset yields a tight approximation of $\mathcal{I}({\bf \bar x};{\bf \bar y}|{\bf \bar H})$:
\begin{align}\label{eq:I hat}
	\hat{\mathcal{I}}({\bf \bar x};{\bf \bar y}|{\bf \bar H})&=2\log_2|\mathcal{S}|-\log_2\sum_{{\bf \bar x}_{i}\in  \mathcal{S}}\sum_{{\bf \bar x}_{j}\in  \mathcal{S}}\exp\left(-\frac{{\|{\bf \bar H}({\bf \bar x}_{i}-{\bf \bar x}_{j})\|^2}}{{2{\sigma}^2}}\right)\nonumber\\
	&=-\log_2\frac{1}{|\mathcal{S}|^2}\sum_{{\bf \bar x}_{i}\in  \mathcal{S}}\sum_{{\bf \bar x}_{j}\in  \mathcal{S}}\exp\left(-\frac{{\|{\bf \bar H}({\bf \bar x}_{i}-{\bf \bar x}_{j})\|^2}}{{2{\sigma}^2}}\right).
\end{align} 
In \eqref{eq:I hat}, the elements of ${\bf \bar H}$ follow the complex Gaussian distribution. Therefore, for given symbol vectors ${\bf \bar x}_i, {\bf \bar x}_j$, $\kappa=\frac{\|{\bf {\bar H}}({\bf {\bar x}}_i-{\bf {\bar x}}_j)\|^2}{2{\sigma}^2}$ follows the Gamma distribution with PDF $f_{\kappa}(w) \sim \Gamma\left(N_{\rm r},\frac{\|{\bf x}_i-{\bf x}_j\|^2}{2{{\sigma}^2}}\right)$. The exception of the mutual information is approximated by applying Jensen's inequality to $\hat{\mathcal{I}}$ as
\begin{align}\label{eq:I expectation}
	\mathbb{E}_{\bf \bar H}&\left[\mathcal{I}\left(\bf \bar x ; \bf \bar y| \bf \bar H \right)\right]\approx\mathbb{E}_{\bf \bar H}\left[\hat{\mathcal{I}}\left(\bf \bar x ; \bf \bar y| \bf \bar H \right)\right] \nonumber\\
	&\geq-\log_2\frac{1}{|\mathcal{S}|^2}\sum_{{\bf \bar x}_{i}\in  \mathcal{S}}\sum_{{\bf \bar x}_{j}\in  \mathcal{S}}\mathbb{E}_{\bf \bar H}\left[\exp\left(-\frac{{\|{\bf \bar H}({\bf \bar x}_{i}-{\bf \bar x}_{j})\|^2}}{{2{\sigma}^2}}\right)\right].
\end{align}
The lower bound is due to the concavity of the log function. In addition, the expectation in \eqref{eq:I expectation} is simply calculated using the moment generating function (MGF) of the Gamma random variable $\kappa$, i.e., $\mathbb{E}\left[e^{-\kappa}\right]=\left(1+\frac{\|{\bf \bar x}_i-{\bf \bar x}_j\|^2}{2{\sigma}^2}\right)^{-N_{\rm r}}$. As a result, the approximate expression of the average mutual information is
\begin{align}\label{eq:I final}
	\mathbb{E}_{\bf \bar H}\left[\mathcal{I}\left(\bf \bar x ; \bf \bar y| \bf \bar H \right)\right]\approx-\log_2\frac{1}{|\mathcal{S}|^2}\sum_{{\bf \bar x}_{i}\in  \mathcal{S}}\sum_{{\bf \bar x}_{j}\in \mathcal{S}}{\left(1+\frac{\|{\bf \bar x}_i-{\bf \bar x}_j\|^2}{2{\sigma}^2}\right)^{-N_{\rm r}}}\nonumber\\
	=2\log_2|\mathcal{S}|-\log_2\sum_{{\bf \bar x}_{i}\in  \mathcal{S}}\sum_{{\bf \bar x}_{j}\in \mathcal{S}}{\left(1+\frac{\|{\bf \bar x}_i-{\bf \bar x}_j\|^2}{2{\sigma}^2}\right)^{-N_{\rm r}}},
\end{align}
which completes the proof.
\end{proof}

Because the proposed SLM uses a set of lattice points with symmetric minimum distance, i.e., $\|{\bf \bar x}_i-{\bf \bar x}_j\|^2\geq d^2_{\rm min}({\bf \Lambda}), ~({\bf \bar x}_i\neq{\bf \bar x}_j)$, we can further simplify the average mutual information in the following corollary:

\vspace{0.1cm}
{\bf Corollary 1:} A lower bound of $\mathbb{E}_{\bf \bar H}\left[\hat{\mathcal{I}}\left(\bf \bar x ; \bf \bar y| \bf \bar H \right)\right]$ for SLM using lattice $\bf \Lambda$ is
\begin{align} \label{eq:corollary}
& \mathbb{E}_{\bf \bar H}\left[\hat{\mathcal{I}}\left(\bf \bar x ; \bf \bar y| \bf \bar H \right)\right]\nonumber\\
 &\geq \log_2|\mathcal{S}|-\log_2\left(1+(|\mathcal{S}|-1)\left(1+\frac{d^2_{\rm min}(\bf \Lambda)}{2{\sigma}^2}\right)^{-N_{\rm r}}\right).
\end{align}
The result in Corollary 1 clearly shows that a higher average mutual information is achievable when dense lattices (or equivalently a large $d^2_{\rm min}({\bf \Lambda})$) are used and the number of receive antennas increases.
\begin{figure}[t]
    \centering
    \epsfig{file=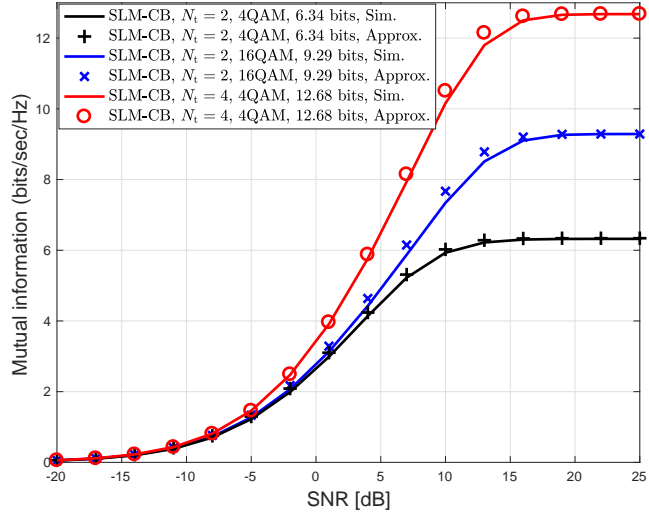, width=10cm}
    \caption{ Simulated (lines) and analytic (symbols) average mutual information results when using the proposed SLM with cubic lattices in $N_{\rm t}$ $\times$ 4 ($N_{\rm t}$ = 2, 4) MIMO systems.}\label{mu_info_analysis}
\end{figure}

Fig. \ref{mu_info_analysis} shows that the proposed approximation of the average mutual information derived in \eqref{eq:I final} agrees well with simulations at all SNR region for MIMO systems with antenna configurations $(N_{\rm t},N_{\rm r},M) = (2,4,2),(2,4,4),$ and $(4,4,2)$, and various modulation sizes.

\subsection{Average Symbol-Vector-Error Probability}
This subsection calculates an upper bound of ASVEP for MIMO systems when the proposed SLM is used.

\vspace{0.1cm}
{\bf Proposition 4:} An upper bound of ASVEP is 
\begin{align} \label{eq:SVE_bounded}
 P_{{\rm ASVEP}}\leq\sum_{{\bf \bar x}_{i}\in  \mathcal{S}}\sum_{{\bf \bar x}_j\neq {\bf \bar x}_i}\frac{{\mu_{ij}}^{N_{\rm r}}\sum_{k=0}^{N_{\rm r}-1}{{N_{\rm r}-1+k}\choose{k}}(1-{\mu_{ij}})^k}{|\mathcal{S}|}.
\end{align}

\begin{proof}

Let $P\left({\bf \bar x}_i\rightarrow{\bf \bar x}_j\right)$ be the pairwise symbol vector error probability (PSVEP) when deciding ${\bf \bar x}_j$ given that ${\bf \bar x}_i$ is transmitted. Using an union bounding technique \cite{TseBook},  ASVEP is upper bounded by the sum of all possible PSVEPs as
\begin{align}\label{eq:SVE1}
P_{{\rm ASVEP}}&\leq{\mathbb{E}_{\bf \bar x}}\left[\sum_{j} P({\bf \bar x}_i\rightarrow{\bf \bar x}_j) \right] \nonumber \\
&=\frac{1}{|\mathcal{S}|}\sum_{{\bf \bar x}_i \in \mathcal{S}} \sum_{{\bf \bar x}_j\neq {\bf \bar x}_i}P({\bf \bar x}_i\rightarrow{\bf \bar x}_j).
\end{align}
By using the ML detection principle in \eqref{eq:ML}, PSVEP conditioned on ${\bf \bar H}$ is given by
\begin{align}\label{eq:SVE3}
P({\bf \bar x}_i\rightarrow{\bf \bar x}_j|{\bf \bar H})&={P}\left(\|{\bf \bar v}+{\bf \bar H}({\bf \bar x}_i-{\bf \bar x}_j)\|^2<\|{\bf \bar v}\|^2\Big|~{\bf \bar H}\right) \nonumber \\
&={P}\left({\rm Re}\left\{{\bf \bar v}^H{\bf \bar H}({\bf \bar x}_i-{\bf \bar x}_j)\right\}<-\frac{1}{2}\|{\bf \bar H}({\bf \bar x}_i-{\bf \bar x}_j)\|^2\Big|~{\bf \bar H}\right) \nonumber \\
&=Q\left(\sqrt{\kappa}\right),
\end{align}
where $Q(x)=\int_{x}^{\infty}\frac{1}{\sqrt{2\pi}}e^{-\frac{t^2}{2}}dt$ denotes the tail probability of a standard Gaussian distribution. Thus, the PSVEP is obtained by computing the integral, i.e.,
\begin{align}\label{eq:SVE4}
P({\bf \bar x}_i\rightarrow{\bf \bar x}_j)&={\mathbb{E}_{\bf \bar H}}\Big[P({\bf \bar x}_i\rightarrow{\bf \bar x}_j\Big|{\bf \bar H})\Big] \nonumber \\
&=\int_{w=0}^{\infty}Q\left(\sqrt{w}\right)f_{\kappa}(w)dw \nonumber\\
&={\mu_{ij}}^{N_{\rm r}}\sum_{k=0}^{N_{\rm r}-1}{{N_{\rm r}-1+k}\choose{k}}(1-{\mu_{ij}})^k,
\end{align}
where ${\mu_{ij}}=\frac{1}{2}\left(1-\sqrt{\frac{\|{\bf \bar x}_i-{\bf \bar x}_j\|^2}{4{\sigma}^2+\|{\bf \bar x}_i-{\bf \bar x}_j\|^2}}\right)$ and the third equality follows from the closed form expression given in \cite{Alouini1999}. Substituting \eqref{eq:SVE4} into \eqref{eq:SVE1} yields
\begin{align}\label{eq:SVE5}
P_{{\rm ASVEP}}\leq\sum_{{\bf \bar x}_{i}\in  \mathcal{S}}\sum_{{\bf \bar x}_j\neq {\bf \bar x}_i}\frac{{\mu_{ij}}^{N_{\rm r}}\sum_{k=0}^{N_{\rm r}-1}{{N_{\rm r}-1+k}\choose{k}}(1-{\mu_{ij}})^k}{|\mathcal{S}|},
\end{align}
which completes the proof. 
\end{proof}

We can further simplify ASVEP in Proposition 4 by using the well known upper bound of the Gaussian Q-function. Using following inequality $Q(x)\leq\frac{1}{2}\exp\left({-\frac{x^2}{2}}\right)$, the SVPEP in \eqref{eq:SVE3} is upper bounded by	
\begin{align}\label{eq:SVE bound 1}
P({\bf \bar x}_i\rightarrow{\bf \bar x}_j|{\bf \bar H})=Q\left(\sqrt{\kappa}\right)\leq\frac{1}{2}\exp\left({-\frac{\kappa}{2}}\right).
\end{align}
By marginalizing with respect to $\kappa$
\begin{align}\label{eq:SVE bound 2}
P({\bf \bar x}_i\rightarrow{\bf \bar x}_j)\leq\frac{1}{2}{\mathbb{E}_{\kappa}}\left[\exp\left({-\frac{\kappa}{2}}\right)\right].
\end{align}
Using the MGF of the Gamma distribution yields 
\begin{align}\label{eq:SVE bound 3}
P({\bf \bar x}_i\rightarrow{\bf \bar x}_j)\leq\frac{1}{2}\left(1+\frac{\|{\bf \bar x}_i-{\bf \bar x}_j\|^2}{4{\sigma}^2}\right)^{-N_{\rm r}}.
\end{align}
Consequently, from \eqref{eq:SVE1}, we obtain the upper bound as
\begin{align} \label{eq:SVE_bounded}
P_{\rm ASVEP}\leq\frac{1}{2|\mathcal{S}|}\sum_{{\bf \bar x}_{i}\in  \mathcal{S}}\sum_{{\bf \bar x}_j\neq {\bf \bar x}_i}\left(1+\frac{\|{\bf \bar x}_i-{\bf \bar x}_j\|^2}{4{\sigma}^2}\right)^{-N_{\rm r}},
\end{align}
which is a similar form of \eqref{eq:I final}.

From the fact that $\|{\bf \bar x}_i-{\bf \bar x}_j\|^2 \geq d^2_{\rm min}({\bf \Lambda}) ({\bf \bar x}_i\neq{\bf \bar x}_j)$, we further simplify \eqref{eq:SVE_bounded} in the following corollary.

\vspace{0.1cm}
{\bf Corollary 2:} A lower bound of ASVEP for SLM using lattice $\bf \Lambda$ is
\begin{align} \label{eq:Corollary2}
P_{\rm ASVEP} &\leq \frac{1}{2|\mathcal{S}|} \sum_{{\bf \bar x}_{i}\in  \mathcal{S}}\sum_{{\bf \bar x}_j\neq {\bf \bar x}_i}\left(1+\frac{ d^2_{\rm min}({\bf \Lambda}) }{4{\sigma}^2}\right)^{-N_{\rm r}} \nonumber \\
&= \frac{(|\mathcal{S}|-1)}{2} \left( 1 + \frac{ d^2_{\rm min}({\bf \Lambda}) }{4{\sigma}^2}\right)^{-N_{\rm r}}.
\end{align}
Similar to Corollary 1, the result in Corollary 2 shows that ASVEP decreases as the minimum distance of the lattice $ d^2_{\rm min}({\bf \Lambda})$ increases or the number of receive antennas $N_{\rm r}$ increases.
\section{Lattice Sphere Decoding}
In this section, we present a low complexity detection method for SLM, referred to as \textit{lattice sphere decoding (LSD)}. One drawback of the SLM introduced in Section III is that the ML detection complexity increases exponentially with the number of transmit antennas or the modulation size, i.e., $\mathcal{O}\left((M+1)^{2N_{\rm t}}\right)$. Therefore, for massive MIMO systems with a large number of transmit antennas, SLM is not appropriate due to the forbidding detection complexity. The proposed LSD overcomes this drawback by reducing the effective search space to the closest lattice vectors from an initially estimated lattice vector.
\subsection{Lattice Sphere Decoding}
This subsection explains the key idea and the algorithm of lattice sphere decoding (LSD). The core idea of LSD is to search only the closest lattice vectors from an initially estimated lattice vector. 

{\bf 1) Initial estimate:} The first step of LSD is to find an initial estimate $\hat{\bf x}$ by using linear detection methods such as a minimum mean-square error (MMSE) detection method, namely,
\begin{align}
	{\bf \hat x}= \left({\bf H}^{\top}{\bf H}+\frac{\sigma^2}{E_s/N_{\rm t}}{\bf I}\right)^{-1}{\bf H}^{\top}{\bf y}.
\end{align}

{\bf 2) Vector quantization:} The estimated $\hat{\bf x}$ is quantized to the closest lattice vector by using the vector quantization function\footnote{For more details on the lattice vector quantization in \cite{Conway1982}, see Appendix A.} $Q:\mathbb{R}^{2 N_{\rm t}}\rightarrow {\bf \Lambda}^{\rm BW}$ in \cite{Conway1982}, i.e.,
\begin{align}
	{\bf \bar x}=Q({\bf \hat x}).
\end{align}
Since the transmitted signal is a lattice vector that satisfies the maximum power constraint $P_{\rm max}$ by the construction of SLM, we need to rescale the initial estimate ${\bf \hat x}$ if $\|Q({\bf \hat x})\|^2> P_{\rm max}$, i.e., 
\begin{align}
	{\bf \bar x}=Q\left(\frac{{\bf \hat x}}{\|{\bf \hat x}\|}\sqrt{P_{\rm max}}\right).
\end{align} 

{\bf 3) Reduced lattice set construction inside a sphere:} We construct a subset $\hat {\mathcal{X}}^{\rm BW}\subset {\bf \Lambda}^{\rm BW}$ whose elements lie in the sphere centered at the quantization output vector ${\bf \bar x}$ with radius of $d$. In particular, we set the radius to the smallest Euclidean norm, i.e., $d=d^2_{\rm min}({\bf \Lambda}^{\rm BW})$. This sphere radius guarantees that the cardinality of $\hat {\mathcal{X}}$ is equal to the kissing number of a lattice plus one, i.e, $|\hat{\mathcal{X}}^{\rm BW}| = K_{\rm min}({\bf \Lambda}^{\rm BW})+1$. 

Using the fact that a lattice is closed under addition, we find the elements of the subset $\hat{ \mathcal{X}}^{\rm BW}$ in a systematic manner. Let define a set $\mathcal{D}_{\rm min}$ that has elements with the smallest Euclidean norm $d^2_{\rm min}({\bf \Lambda})$, i.e., the set with the closest lattice vectors from origin. By the lattice property, the cardinality of $\mathcal{D}_{\rm min}$ is equal to the kissing number of a lattice, i.e, $|\mathcal{D}_{\rm min}| = K_{\rm min}({\bf \Lambda})$. Using $\mathcal{D}_{\rm min}$, it is possible to find the elements of $\hat {\mathcal{X}}^{\rm BW}$ by adding the initial estimate lattice point ${ \bf \bar x}$ to the elements in $\mathcal{D}_{\rm min}$, namely,
\begin{align}
\hat {\mathcal{X}}^{\rm BW} = \{ {\hat {\bf {x}}}|{\hat {\bf x}} = \bar {\bf x} + {\bf d}, {\bf d} \in \mathcal{D}_{\rm min} \cup {\{ {\bf 0}\}}\}.
\end{align}
Under the maximum power constraint $P_{\rm max}$, the set $\hat {\mathcal{X}}^{\rm BW}$ is further reduced to a set $\mathcal{X}^{\rm BW}$ by only merging the vectors which satisfy this power constraint.
\begin{align}
	\mathcal{X}^{\rm BW} = \{ {\bf x} | {\bf x} \in \hat {\mathcal{X}}^{\rm BW} ,||{\bf x}||^2 \leq P_{\rm max}\}.
\end{align}

{\bf 4) Decoding over the reduced set:} Using the reduced lattice set in the sphere, $\mathcal{X}^{\rm BW}$, the receiver performs ML detection over the reduced set, i.e., 
\begin{align}
	{\hat {\bf s}} = \arg \min_{{\bf x} \in \mathcal{X}^{\rm BW}}||{\bf y}-{\bf H}{\bf x}||^2.
\end{align}

\begin{figure}[t]
	\centering
  \includegraphics[width=0.5\textwidth]{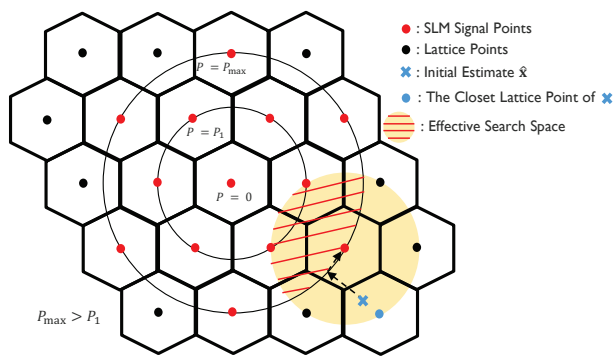}\\
  \caption{The illustration of the proposed lattice sphere decoding (LSD) for SLM.} \label{LSD_fig}
\end{figure}

As seen in Fig. \ref{LSD_fig}, the blue x-mark represents the initial estimate $\hat {\bf x}$. This blue x-mark is quantized to the lattice vector (the blue circle). Because the power of the blue circle exceeds $P_{\rm max}$, the blue x-mark are rescaled and re-quantized. Then, we obtain the lattice set inside the sphere (the red stripes region), which includes the quantized lattice vector. Finally, we calculate ML metrics over the reduced lattice set $\mathcal{X}^{\rm BW}$. The proposed algorithm is summarized in Algorithm 2. 
\begin{algorithm}[t]\small
    \caption{Lattice Sphere Decoding.}
  \begin{algorithmic}[1]
    \INPUT Received signal ${\bf y}$, Channel matrix ${\bf H}$, SLM signal set $\mathcal{S}^{\rm BW}(N_{\rm t},P_{\rm max})$.
    \OUTPUT Estimated input signal point $\hat {\bf s} \in \mathcal{S}^{\rm BW}(N_{\rm t},P_{\rm max})$.
      \STATE Using MMSE estimator, calculate the initial estimate ${\hat {\bf x}}$. 
      \STATEx ${\hat {\bf x}} = [{\hat x}_1, {\hat x}_2, \cdots, {\hat x}_{2N_{\rm t}}] = ({\bf H}^{\rm T} {\bf H} + \frac{\sigma^2}{E_{\rm s}/ N_{\rm t}} {\bf I})^{\rm T} {\bf y}$.
            	\STATEx If the power of $Q({\hat {\bf x}})$ exceeds  $P_{\rm max}$, rescale the power of ${\hat {\bf x}}$ to $P_{\rm max}$ and then quantize it.
            	\IF{$||Q({\hat {\bf x}})||^2 >
            	 P_{\rm max}$}
            	 \STATE $\hat {\bf x} = \frac{\hat {\bf x}} {||\hat {\bf x}||} \cdot \sqrt{P_{\rm max}}$.
            	\ENDIF
            	\STATE  ${\bar {\bf x}}= Q(\hat {\bf x})$. 
			\STATEx Obtain reduced lattice sphere set $\hat {\mathcal{X}}^{\rm BW}$.
     		\STATE $\hat {\mathcal{X}}^{\rm BW} = \{ {\hat {\bf {x}}}|{\hat {\bf x}} = \bar {\bf x} + {\bf d}, {\bf d} \in \mathcal{D}_{\rm min} \cup {\{{\bf 0}\}}\}$.
     		\STATEx Merge the elements that satisfy the maximum power constraint $P_{\rm max}$ into $\mathcal{X}^{\rm BW}$.
     		\vspace{0.05cm}
     		\STATE $\mathcal{X}^{\rm BW} = \{ {\bf x} | {\bf x} \in \hat {\mathcal{X}}^{\rm BW} ,||{\bf x}||^2 \leq P_{\rm max}\}.$
      \STATE ${\hat {\bf s}} = \arg \min_{{\bf x} \in \mathcal{X}^{\rm BW}}||{\bf y}-{\bf H}{\bf x}||^2$.
  \end{algorithmic}
\end{algorithm}

Throughout this section, we mainly focus on LSD with Barnes-Wall lattices. With a similar manner, we are able to extend the LSD algorithm for cubic lattices. In the case of cubic lattices, we use a simple element-wise quantization function in which every  elements of the estimated lattice vector is sliced to the closest integer of $\mathcal{\bar C}=\left\{\frac{-M}{2},\frac{-M}{2}+1,\ldots,0,\ldots,\frac{M}{2}-1,\frac{M}{2}\right\}$. Also, the power constraint is changed into the $M$-ary constellation constraint.  

To provide a better understanding of the proposed LSD, it is instructive to consider an example. 

\vspace{0.1cm}
{\bf Example 3:} Consider a system in which $N_{\rm t} = 2$ and the signal points in Table II are used for SLM-BW. In other words, SLM-BW is designed with the maximum power constraint of 6, i.e., $P_{\rm max} = 6$.
Following by Algorithm 2, we firstly obtain an initial estimate $\hat {\bf x}$. For example, 
\begin{align}
	\hat {\bf x} = \left[ +1.32,~ -2.51, ~ -0.41, ~ +2.70 \right]^{\rm T}.
\end{align}
Then, $\hat {\bf x}$ is quantized to the closest lattice vector $Q(\hat {\bf x})$\footnote{In Appendix A, we provide the quantization procedures of $\hat {\bf x}$ as an example.}.
\begin{align}
		Q(\hat {\bf x}) = \left[ +1,~ -2, ~ 0, ~ +3 \right]^{\rm T}.
\end{align}
	The power of $Q(\hat {\bf x})$ equals $14$. Therefore, we rescale the power of $\hat {\bf x}$ to $P_{\rm max}$.
	\begin{align}
		&\frac{{\bf \hat x}}{\|{\bf \hat x}\|}\sqrt{P_{\rm max}} = \left[ +0.82,~ -1.57, ~ -0.25, ~ +1.68 \right]^{\rm T}. \nonumber
		\\
		&\bar {\bf x} = Q\left(\frac{{\bf \hat x}}{\|{\bf \hat x}\|}\sqrt{P_{\rm max}}\right) = \left[ +1,~ -1, ~ 0, ~ +2\right]^{\rm T}.
	\end{align}	
After obtaining the quantized output $\bar {\bf x}$, the lattice sphere set $\mathcal{X}^{\rm BW}$ is generated. $\mathcal{X}^{\rm BW}$ of SLM-BW is obtained by adding $\bar {\bf x}$ to $\mathcal{D}_{\rm min}(= \mathcal{S}^{2} ~\text{in Table II})$ and removing the elements that violate the power constraint as follows:
\begin{align}
{\mathcal{X}^{\rm BW}} =
	\left\{ \left[ \begin{array}{c} +1 \\ -1 \\ 0 \\ +2 \end{array} \right] , \left[ \begin{array}{c} 0 \\ 0 \\ 0 \\ +2 \end{array} \right], \left[ \begin{array}{c} 0 \\ -1 \\ -1 \\ +2 \end{array} \right], \cdots , \left[ \begin{array}{c} +1 \\ -2 \\ 0 \\ + 1 \end{array} \right] , \left[ \begin{array}{c} +1 \\ -1 \\ -1\\ +1 \end{array} \right] \right\}. 
\end{align}
Then, the receiver calculates the ML metrics over the twelve symbol vectors in $\mathcal{X}^{\rm BW}$ and decides a symbol vector that has the minimum value of $\|{\bf y}-{\bf H}{\bf x}\|^2$. Similarly, the receiver performs the same procedure for SLM-CB.

\subsection{Detection Complexity Analysis}
In this subsection, the detection complexities of the proposed LSD and ML detectors for SLM are analyzed. In addition, these are compared to the complexities of detectors for the existing spatial modulation and the spatial multiplexing method. Each real addition, multiplication, and rounding operation are counted as one floating operation (flop).

\vspace{0.1cm}
\textbf{1) MLD:} Let $N_{\rm a}$ be the the number of activated dimensions in an $2N_{\rm t}$-dimensional real space. Therefore, $N_{\rm a}$ of SM equals 2, $N_{\rm a}$ of spatial multiplexing equals $2N_{\rm t}$, and $N_{\rm a}$ of SLM has values between 0 and $2N_{\rm t}$. The complexity of ML detector in \eqref{eq:ML} can be computed as follows:
\begin{itemize}
\item The operator ${\bf H}{\bf x}$ requires $N_{\rm r} N_{\rm a}$ real multiplications and $N_{\rm r} (N_{\rm a}-1)$ real additions.
\item The operator ${\bf y-Hx}$ requires $N_{\rm r}$ real additions.
\item The operator $\| {\bf y-Hx} \|^2$ requires $N_{\rm r}$ real multiplications and $N_{\rm r}-1$ real additions.
\end{itemize}
To compute $\|{\bf y-Hx}\|^2_2$, $2N_{\rm r}(N_{\rm a}+1)-1$ flops are required for each symbol vector ${\bf x} \in \mathcal{S}$. SM and spatial multiplexing requires $(4N_{\rm r}-1) \log_2(N_{\rm t} M^2 ) $ flops and $(2N_{\rm r}(N_{\rm t}+1)-1) \log_2(M^{2{N_{\rm t}}})$ flops, respectively. For SLM using cubic lattices, $N_{\rm a}$ varies from 0 to $2N_{\rm t}$; thereby, the required flops of the SML are given by
\begin{align}
	\sum_{N_{\rm a}=0}^{2N_{\rm t}} |\mathcal{S}^{N_{\rm a}}| (2N_{\rm r}(N_{\rm a}+1)-1),
\end{align}
where $|\mathcal{S}^{N_{\rm a}}|= {2N_{\rm t} \choose N_{\rm a}} M ^{N_{\rm a}}$. Note that the detection complexity of SLM with barnes-wall lattices is similar to that of SLM using cubic lattices.

\vspace{0.1cm}
\textbf{2) LSD:} As described in Algorithm 2, the effective search space of LSD is determined by the cardinality of $\mathcal{X}$. This set size can be changed depending on the lattice structure for SLM and the quantized method. For convenience, we analyze the worst case of LSD, i.e, $|\mathcal{X}| = |\mathcal{D}_{\rm min}| + 1$. The complexity of LSD in Algorithm 2 is computed as follows:
\begin{itemize}
	\item By following \cite{Xiao2017}, the MMSE estimator $({\bf H}^{\top} {\bf H} + \frac{\sigma^2}{E_{\rm s}/ N_{\rm t}} {\bf I})^{\top} {\bf y}$ requires $2N_{\rm t}^3+6 N_{\rm t}^2 N_{\rm r} + 6 N_{\rm t}^2 + 4 N_{\rm t} N_{\rm r}$ real multiplications and $2N_{\rm t}^3 + 6N_{\rm t}^2N_{\rm r} + N_{\rm t}^2 + 2N_{\rm t}N_{\rm r}$ real additions.
	\item As indicated in \cite{Conway1982}, the quantization complexity order of SLM-BW with $N_{\rm t}$ antennas follows $\mathcal{O}\left(2N_{\rm t}\right)$. For example, SLM-BW with $N_{\rm t} = 2$ requires $8N_{\rm t}-2$ flops.
	\item The construction of $\mathcal{X}^{\rm BW}$ requires $|\mathcal{D}_{\rm min}|(4N_{\rm t} - 1)$ real additions and $2 |\mathcal{D}_{\rm min}|N_{\rm t}$ real multiplications.
	\item The operator $\| {\bf y-Hx} \|^2$ is computed $| \mathcal{D_{\rm min}} | + 1$ times. Therefore, the entire flops required are $(|\mathcal{D_{\rm min}}| + 1) (2N_{\rm r}(2N_{\rm t}+1)-1)$. 
\end{itemize}

We summarize the detection complexities of ML, MMSE, and LSD in Table IV and compared with other modulation and detection methods.

\begin{table}[t]
\centering
	\begingroup 
	\caption{Detection Complexities of Various Modulation Methods.}
	\resizebox{\columnwidth}{!}{
\begin{tabular}{||c|c||c|c||c|c||}
\Xhline{3\arrayrulewidth}
\multicolumn{2}{||c||}{Spatial Lattice Modulation} & \multicolumn{2}{c||}{Spatial Modulation} & \multicolumn{2}{|c||}{Spatial Multiplexing} \\
\Xhline{3\arrayrulewidth}
Detection & \# of real-valued flops & Detection & \# of real-valued flops & Detection & \# of real-valued flops \\ \hline
ML & $\sum_{N_{\rm a}=0}^{2N_{\rm t}} {2N_{\rm t} \choose N_{\rm a}} M^{N_{\rm a}} (2N_{\rm r}(N_{\rm a}+1)-1)$ & ML & $(4N_{\rm r}-1) \log_2(N_{\rm t} M^2 ) $ & ML & $(2N_{\rm r}(N_{\rm t}+1)-1) \log_2(M^{2{N_{\rm t}}})$ \\ \hline
$\begin{array}{c}
	\text{Proposed LSD} \\ \text{(worst case)} \end{array}$ & $\begin{array}{c}
4N_{\rm t}^{3}+12 N_{\rm t}^2 N_{\rm r}  + 7 N_{\rm t}^2 + 6 N_{\rm t} N_{\rm r} - 4N_{\rm t} \\ |\mathcal{D}_{\rm min}|(4N_{\rm t}N_{\rm r} + 4 N_{\rm t} + 2N_{\rm r} - 1) + \mathcal{O}(2N_{\rm t}) \end{array}$ &  MRRC \cite{Mesleh2008} & $8N_{\rm t} N_{\rm r} + 2$ & MMSE & $\begin{array}{c}
4N_{\rm t}^{3}+12 N_{\rm t}^2 N_{\rm r}  + 7 N_{\rm t}^2  \\ + 6 N_{\rm t} N_{\rm r} +2N_{\rm t} \end{array}$ \\ \hline
\Xhline{3\arrayrulewidth}
\end{tabular}} \label{table:LSD_complexity}
\endgroup
\end{table}
\section{Numerical Results}
This section provides numerical results on the average mutual information, uncoded symbol-vector-error-rate (SVER), and coded frame-error-rate (FER) of the proposed SLM methods. These metrics are used to show SNR gains compared to the existing SM and the spatial multiplexing techniques. All results are obtained by using Monte Carlo simulations on independent flat-fading channel realizations for various spectral efficiencies as a function of SNR.

\subsection{Average Mutual Information}
%

Fig. \ref{2_4_mu_info_comparison} depicts the achievable spectral efficiencies of various transmission strategies for a MIMO system with $N_{\rm t}=2$, $N_{\rm r}=4$, and the $M$-PAM input constellation set per in-phase and quadrature component (equivalently, $M^2$-QAM). The solid-black line shows the capacity of $2\times4$ MIMO channel when using Gaussian input signaling, which serves an upper bound of the spectral efficiencies attained by the other transmission methods. In Fig. \ref{2_4_mu_info_comparison}, the proposed SLM method with cubic lattices (SLM-CB) uses $4$-PAM per dimension, which is able to generates $5^4$ lattice symbol vectors. In case of SLM with Barnes-Wall lattices (SLM-BW), we select $5^4$ elements from $\mathcal{S}^{\rm BW}(2,18)$ in ascending order of the power consumption. The proposed SLM-CB and SLM-BW both achieve the spectral efficiency of $\log_2(5^4)\approx 9.29$ bits/sec/Hz beyond 20 dB SNR. Especially, SLM-BW achieves higher spectral efficiency than that of SLM-CB in the mid SNR because Barnes-Wall lattices provide a higher nominal coding gain than do cubic lattices, $\gamma({\bf \Lambda}_4^{\rm BW}) >\gamma({\bf \Lambda}_4^{\rm CB})$. In contrast, spatial multiplexing and SM achieve the spectral efficiencies of 8 bits/sec/Hz and 5 bits/sec/Hz respectively when SNR is high enough. Similar results are observed when $2$-ary PAM input signal is used. Thus, SLM achieves the higher spectral efficiency than do the conventional methods.

\subsection{Symbol Vector Error Rate}
\begin{figure}[t]
    \centering
    \epsfig{file=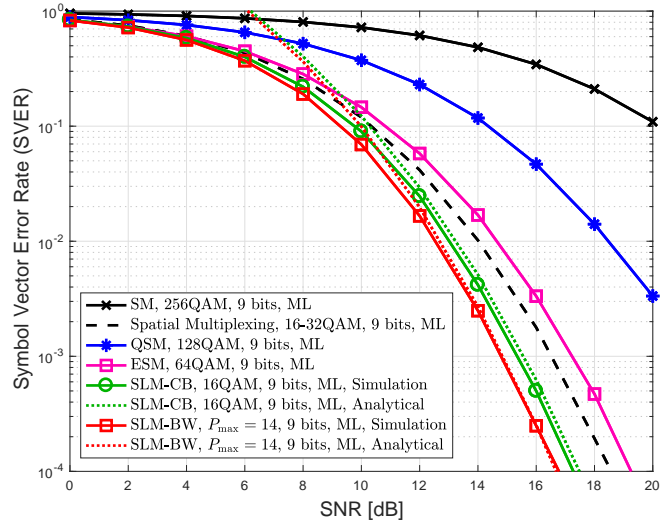, width=10cm}
    \caption{SVERs for the 2$\times$8 MIMO system with the target spectral efficiency of 9 bits/sec/Hz.} \label{2_8_mimo_9bits}
\end{figure}

\begin{figure}[t]
    \centering
    \epsfig{file=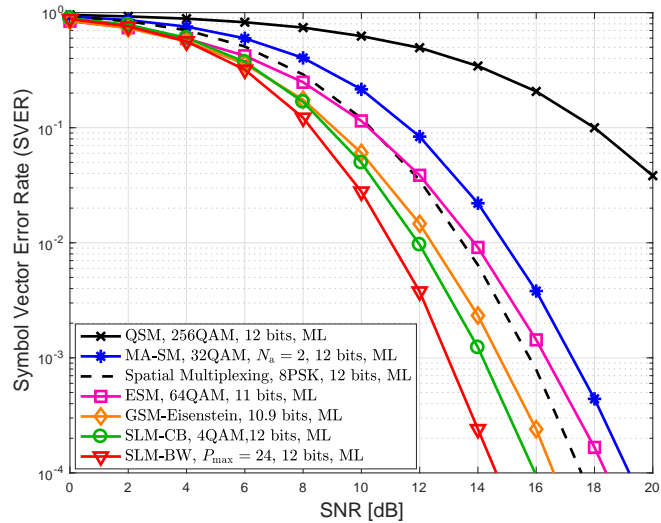, width=10cm}
    \caption{SVERs for the 4$\times$8 MIMO system with the target spectral efficiency of 12 bits/sec/Hz.}
     \label{4_8_mimo_12bits}
\end{figure}

We compare the SVERs of the proposed SLM methods with that of the existing SM, GSM, GSM-Eisenstein, ESM, and spatial multiplexing methods. For fair comparisons, each modulation scheme is allowed to use a different modulation size $M$ to achieve similar target spectral efficiencies. It should be noted that the results of Fig. 5 and Fig. 6 are obtained under ML detection and the results of Fig. 7 and Fig. 8 are obtained under ML detection and LSD. 

Fig. \ref{2_8_mimo_9bits} shows the SVERs when the target spectral efficiency is 9 bits/sec/Hz for the $2\times 8$ MIMO system. To meet this criterion, we set different modulation size for each method:

\begin{itemize}
	\item SM with 256-QAM, which uses 256-QAM to modulate 8 bits and the antenna index to modulate 1 bit;
	\item Spatial multiplexing with 16-QAM and 32-QAM, which uses 16-QAM to modulate 4 bits and 32-QAM to modulate 5 bits, and sends two symbols simultaneously via different transmit antennas;
	\item ESM with 64-QAM, which uses 64-QAM as a primary signal constellation set to modulate 9 bits;
	\item QSM with 128-QAM, which uses 128-QAM to modulate 7 bits and the antenna indies to modulate 2 bit;
	\item SLM-CB with 16-QAM, which is constructed by selecting $512$ elements from $\mathcal{S}^{\rm CB}(2,4)$ in ascending order of the power consumption to modulate 9 bits;
	\item SLM-BW, which is constructed by selecting $512$ elements from $\mathcal{S}^{\rm BW}(2,14)$ in ascending order of the power consumption to modulate 9 bits;
	\end{itemize}

Fig. \ref{2_8_mimo_9bits} shows that the proposed SLM-CB and SLM-BW achieve lower SVERs than all other transmission methods at all SNR. SLM-CB provides about 1 dB SNR gain over the spatial multiplexing at SVER = $10^{-3}$. SLM-BW provides 1 dB SNR gain over SLM-CB. This increase occurs because the Barnes-Wall lattice has a larger nominal coding gain than does cubic lattice. In addition, our analytic expression of SVER obtained from \eqref{eq:SVE5} is shown to be tight at high SNR.

In Fig. \ref{4_8_mimo_12bits}, SVERs are also measured at the target spectral efficiency of 12 bits/sec/Hz for the $4\times 8$ MIMO system. For GSM-Eisenstein and ESM, the target spectral efficiency of 12 bits/sec/Hz is not well defined in \cite{Cheng2015} and \cite{Freudenberger2017}. Alternatively, we choose the target spectral efficiency of GSM-Eisenstein at 10.9 bits/sec/Hz and ESM at 11 bits/sec/Hz. The modulation size $M$ for each method and the corresponding spectral efficiency are summarized in the legend of Fig. \ref{4_8_mimo_12bits}. As a result, SLM-BW and SLM-CB achieve a lower SVER than that of the existing methods. In particular, SLM-BW provides about 3 dB SNR gains over spatial multiplexing at SVER = $10^{-3}$, which is the 1 dB additional gain compared to the $2\times 8$ MIMO system. These gains are obtained because SLM achieves a high spectral efficiency with a low modulation size when the number of transmit antennas is large enough. For example, with $N_{\rm t} = 4 $ SLM using 2-PAM per dimension (or 4-QAM per two dimensions) is able to achieve the spectral efficiency of 12 bits/sec/Hz, whereas the spatial multiplexing method needs to use 8-PSK to attain 12 bits/sec/Hz. Intuitively, the use of low modulation size leads to the increase of $d_{\rm min}^2$ under the transmit power constraint.

\begin{figure}[t]
\centering
\subfigure[\label{fig:1a} \hspace*{-5mm}]{\includegraphics[width=0.3\textwidth]{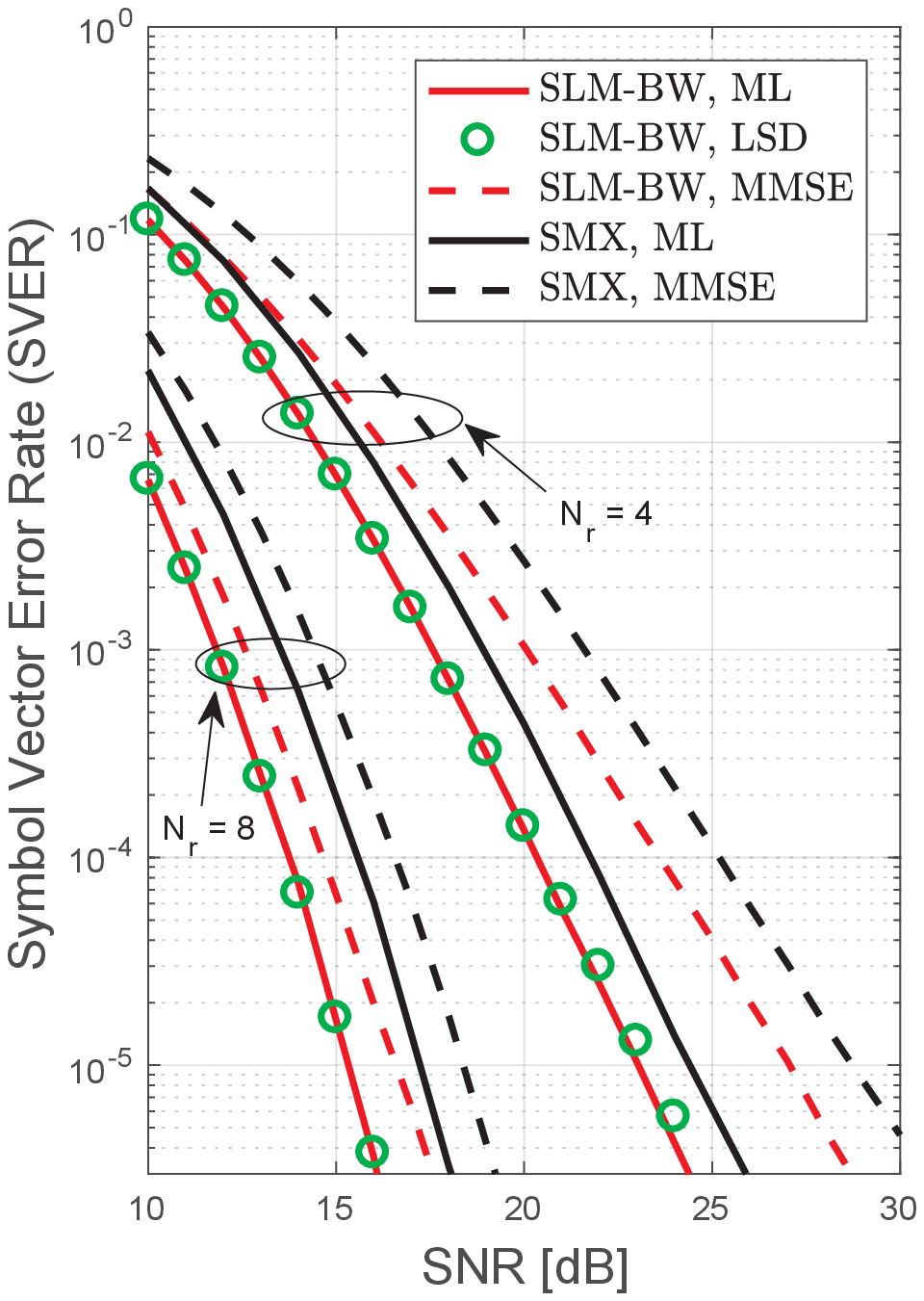}}
\subfigure[\label{fig:1b} \hspace*{-5mm}]{\includegraphics[width=0.3\textwidth]{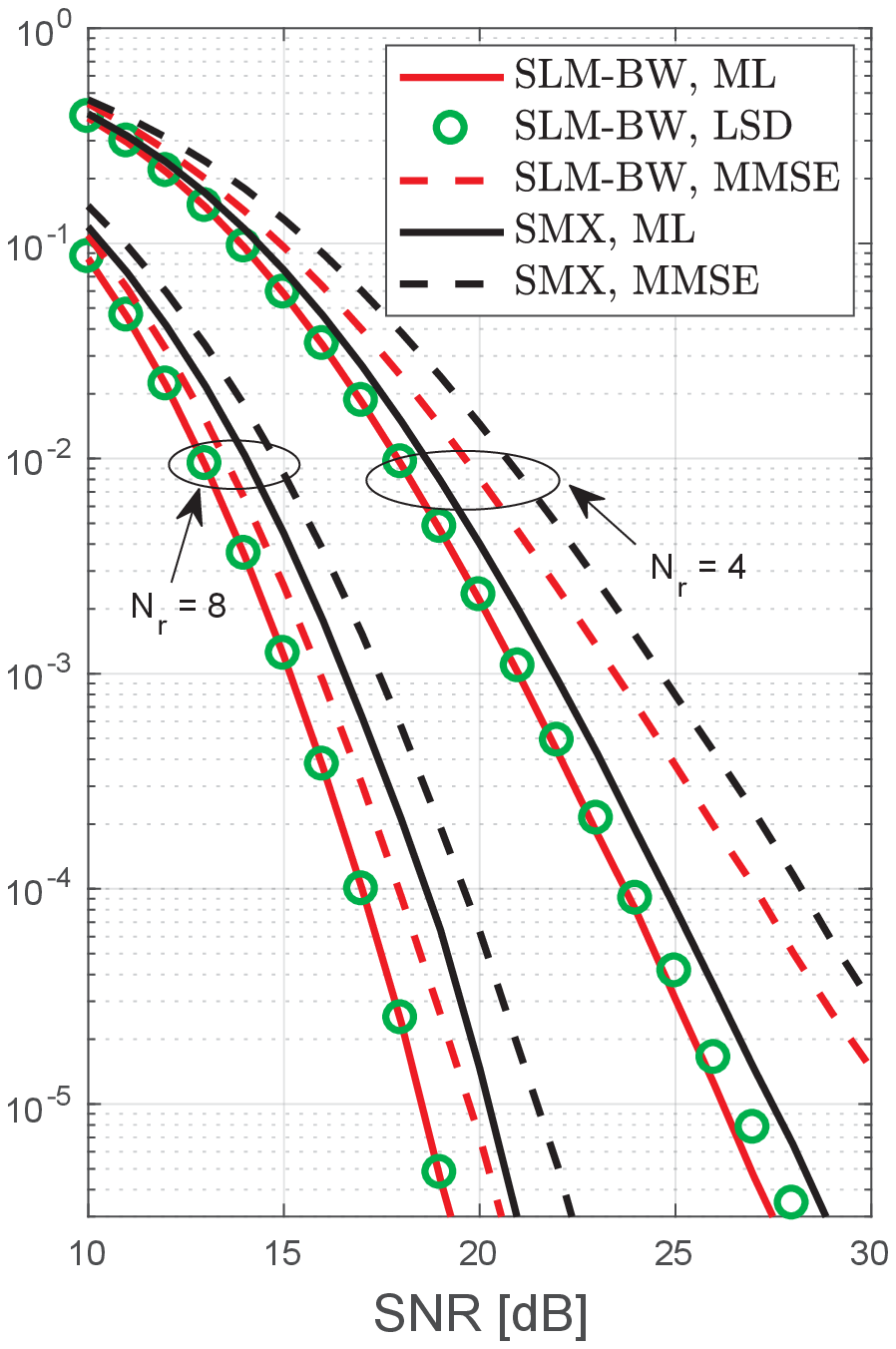}} 
\caption{Comparison of the SVERs under LSD and ML detection with the $2 \times N_{\rm r} ~ (N_{\rm r} =  4, 8)$ MIMO systems. (a): SLM-BW with $\mathcal{S}^{\rm BW}(2,6)$, 7.17 bits/sec/Hz, and spatial multiplexing with 8-PSK and 16-QAM, 7 bits/sec/Hz. (b): SLM-BW with $\mathcal{S}^{\rm BW}(2,14)$, 9.32 bits/sec/Hz, and spatial multiplexing with 16-QAM and 32-QAM, 9 bits/sec/Hz.} \label{LSD_BW}
\end{figure} 

Fig. \ref{LSD_BW} demonstrates the detection performance of the proposed LSD. The proposed LSD is compared to ML and MMSE detection methods. For comparison, we consider $2 \times N_{\rm r} ~ (N_{\rm r} =  4, 8)$ MIMO systems with target spectral efficiencies of about 7 bits/sec/Hz (Fig. \ref{LSD_BW}(a)) and 9 bits/sec/Hz (Fig. \ref{LSD_BW}(b)). The corresponding simulation setups are listed in each sub-caption. For LSD, we use the set $\mathcal{S}^2$ in Table II as $D_{\rm min}$. Therefore, among 145 (Fig. \ref{LSD_BW}(a)), 601 (Fig. \ref{LSD_BW}(b)) SLM vectors, the maximum 25 lattice vectors are computed for the ML metric. In other words, the effective search space of LSD is reduced by $17\%$, $4\%$ of ML detection. Even with the significant complexity reduction, remarkably, the performance of LSD (the green marks) is almost same with SLM with ML detection (the red solid lines). Additionally, SLM with LSD achieves a lower SVER than that of spatial multiplexing with ML and MMSE detection.

\subsection{Frame Error Rate}
\begin{figure}[t]
    \centering
    \epsfig{file=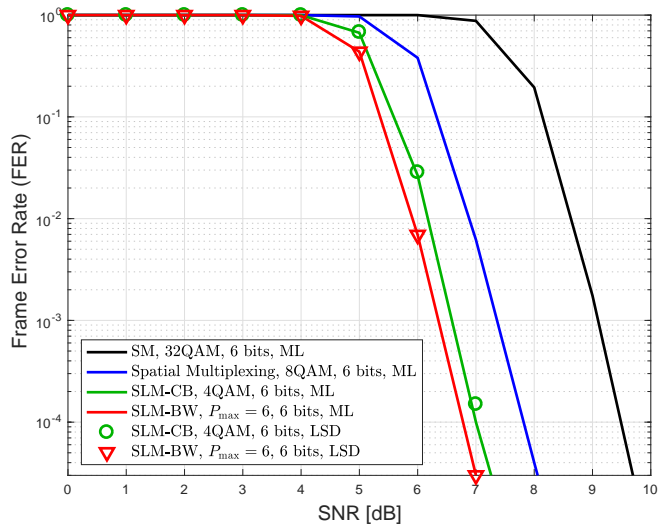, width=10cm}
    \caption{Comparison of the FERs for the 2$\times$4 coded MIMO system.}
         \label{Coded_2_4_mimo_6bits}
\end{figure} 


Fig. \ref{Coded_2_4_mimo_6bits} shows FERs of different transmission strategies in a $2\times4$ coded MIMO system with the target spectral efficiency of 6 bits/sec/Hz. In the simulation, one frame is assumed to consist of 1200 coded bits with a rate of 1/2 turbo code. Thus, 600 information bits are sent per frame. We assume an independent fading channel model in which channel matrices ${\bf H}$ vary at every channel use. Rate-1/2 turbo codes based on parallel concatenated codes with feedforward and feedback polynomial (15,13) in octal notation are used for the simulations. Using soft information, turbo decoding iterates six times, which is sufficient to reduce coded FER compared to the uncoded case. 

For the target spectral efficiency of 6 bits/sec/Hz, SM uses 32QAM, spatial multiplexing uses 8PSK, SLM-CB uses 2-PAM, and SLM-BW is designed with $P_{\rm max} = 6$. Among 81 (SLM-CB), 145 (SLM-BW) lattice vectors, we select 64 lattice vectors in ascending order of the power consumption. In other words, the indices from 1 to 64 in Table I and II are used to modulate 6 bits. For bits-to-symbol mapping, the binary switching algorithm \footnote{The binary switching algorithm in \cite{Zeger1990} provides locally optimal solutions for codebook indexing. Therefore, the performance of SLM in Fig. \ref{Coded_2_4_mimo_6bits} can be further improved by using an novel indexing algorithm.} in \cite{Zeger1990} is applied as a lattice indexing algorithm. We uploaded bits-to-symbol mapping tables in \cite{Choi_Table2017}. Fig. \ref{Coded_2_4_mimo_6bits} demonstrates that the proposed SLM methods provides 1 dB SNR gain over the spatial multiplexing, and 3 dB over SM at FER = $10^{-2}$ when ML detection is applied. The trend of this result is similar to that in Fig. \ref{2_8_mimo_9bits}, in which the number of transmit antennas is the same. The red and green marks in Fig. \ref{Coded_2_4_mimo_6bits} represent the FERs of SLM-BW and SLM-CB when LSD is applied. Soft information for LSD is extracted by calculating log likelihood-ratio values of the vectors in $\mathcal{X}$. Similar to Fig \ref{LSD_BW}, the performance of LSD achieves near ML performance.

\section{Conclusion}
We have presented SLM, a new spatial modulation technique for MIMO systems. By jointly mapping information bits into a set of $2N_{\rm t}$-dimensional lattice points, we have shown that SLM is able to achieve the maximum spectral efficiency at high SNR under the PAM input constraint per dimension. We also demonstrated that the SLM that uses dense Barnes-Wall lattices offers a considerable SNR gain. We derived a tight approximation of the average mutual information and the upper bound of average symbol-vector-error-probability and used simulations to validate the effectiveness of our analysis. In addition, lattice sphere decoding for SLM was proposed to diminish the detection complexity at the receiver. Simulations showed that the performance of LSD closely matches that of ML at every SNR region with practical MIMO setups. 


A promising direction for future work includes a study of impact on SLM for the case where the number of RF chains is greater than the number of transmit antennas, i.e., $N_{\rm t} > N_{\rm RF}$. Other interesting research directions are to devise SLM techniques combined with index modulation methods \cite{Basar2013} for multi-carrier MIMO systems and to devise SLM techniques for the integer-forcing framework in \cite{Nazer2014}.
\appendix
\subsection{Lattice Vector Quantizer}
In this Appendix, we explain the lattice vector quantizers (LVQ) proposed in \cite{Conway1982}. The LVQs in \cite{Conway1982} can quantize any $n$-dimensional vectors into $D_n$ $(D_4 = {\bf \Lambda^{\rm BW}_4})$ or $E_n$ $(E_8 = {\bf \Lambda^{\rm BW}_8})$ lattices. The main advantages of these LVQs are that they can be implemented with simple arithmetic operations and the quantization complexities follow the order of $\mathcal{O} \left( n \right)$. Each lattice requires a different quantization procedure. Due to the paper limitation, we only focus on the LVQ of $D_n$ lattices.


Let $f({\hat x}_i)$ be the closest integer of ${\hat x}_i$. Then, for ${\hat {\bf x}} = [{\hat x}_1,{\hat x}_2,\cdots,{\hat x}_n]^{\rm T} \in \mathbb{R}^n $, we define $f({\hat {\bf x}})$ as 
\begin{align}
		f({\hat {\bf x}}) = [f({\hat x}_1),f({\hat x}_2),\cdots,f({\hat x}_n)]^{\rm T}.
\end{align}
We also define $\delta( {\hat {\bf x}})$ as the difference between ${\hat {\bf x}}$ and $f(\hat{ {\bf x}})$. 
\begin{align}
	\delta( {\hat {\bf x}}) = \left[{\hat x}_1-f({\hat x}_1),{\hat x}_2-f({\hat x}_2),\cdots,{\hat x}_n-f({\hat x}_n) \right]^{\rm T}.
\end{align} 
Using $f({\hat{\bf x}})$ and $\delta({\hat {\bf x}})$, we define $g({\hat {\bf x}})$ as 
\begin{align}
	g({\hat {\bf x}}) = [f({\hat x}_1),\cdots,w({\hat x}_k),\cdots,f({\hat x}_n)]^{\rm T},
\end{align} 
which is the same with $f({\hat {\bf x}})$ except the $k$-th element $w({\hat x}_k)$, where $k = \arg\max(|\delta({\hat {\bf x}})|)$.
The function $w({\hat x}_k)$ rounds ${\hat x}_k$ as follows:
\begin{align}
	w({\hat x}_k) &= f({{\hat x}_k}) + 1, \text{ if $\delta({\hat x}_k) \geq 0 $}, \nonumber \\
		   &= f({{\hat x}_k}) - 1, \text{ if $\delta({\hat x}_k) < 0 $}.
\end{align}

The notable fact is that the sum of all $D_n$ lattice vectors should be even, i.e., $mod(\sum_i {\hat x}_i,2) = 0 $. Therefore, we need to check whether $\sum_i f({\hat x_i})$ or $\sum_i g({\hat x}_i)$ is even. Then, we admit the vector which has even sum as the quantized output $Q(\hat {\bf x})$.

\vspace{0.1cm}
{\bf Example 4:} Suppose ${\hat {\bf x}}= [ +1.32,~ -2.51, ~ -0.41, ~ +2.70]^{\rm T} $. Then $f({\hat {\bf x}})$, $\delta{({\hat {\bf x}})}$, and $g({\hat {\bf x}})$ are computed as
\begin{align}
	f({\hat {\bf x}}) &= [+1,~ -3,~ 0,~ +3]^{\rm T}, \nonumber \\
	\delta(\hat {{\bf x}}) &= [+0.32,~ +0.49,~ -0.41,~ -0.30]^{\rm T}, \nonumber \\
	g(\hat {{\bf x}}) &= [+1,~ -2,~ 0,~ +3]^{\rm T}.  
\end{align}
Considering that $\sum_i{f({\hat x}_i)} = 1$ (odd) and $\sum_i{g({\hat x}_i)} = 2$ (even), we select $g({\hat {\bf x}})$ as the quantized output of $\hat {\bf x}$, i.e., $Q({\hat {\bf x}}) = g({\hat {\bf x}})$. 


%


\end{document}